\documentclass[aps,pre,twocolumn,preprint,groupedaddress,amsmath,amssymb,showpacs]{revtex4-1}

\usepackage{graphicx}% Include figure files
\usepackage{dcolumn}% Align table columns on decimal point
\usepackage{bm}% bold math　　
\usepackage{color} 
\usepackage{amsmath}    % need for subequations
\usepackage{subfigure}

\newcommand{\lam}{\lambda}

\newcommand{\largep}{{\bf p}}

\newcommand{\ep}{\varepsilon}

\newcommand{\largeq}{{\bf q}}

\begin{document}

\title{Quantum and classical chaos of a two-electron system in a quantum wire
}

\author{Shumpei Masuda}
\email[]{syunpei710@cmpt.phys.tohoku.ac.jp}
\affiliation{Department of Physics,
Tohoku University,
Sendai 980, Japan}
\author{Shin-ichi Sawada}
\affiliation{Department of Physics, Kwansei Gakuin University, Gakuen 2-1, 
Sanda 669-1337, Japan}
\author{Yasushi Shimizu}
\affiliation{Department of Physics, Ritsumeikan University, Noji-higashi 1-1-1, 
Kusatsu 525-8577, Japan}

\date{\today}

\begin{abstract}
We study classical and quantum dynamics of two spinless particles confined
in a quantum wire with repulsive or attractive Coulomb interaction. The
interaction induces irregular dynamics in classical mechanics, which
reflects on the quantum properties of the system in the energy level
statistics (the signatures of quantum chaos). We investigate especially
closer correspondence between the classical and quantum chaos. The present
classical dynamics has some scaling property, which the quantum counterpart
does not have. However, we demonstrate that the energy level statistics
implies the existence of the corresponding scaling property even in the
quantum system. Instead of ordinary maximum Lyapunov exponent (MLE), we
introduce a novel kind of MLE, which is shown to be suitable measure of
chaotic irregularity for the present classical system. We show that tendency
of the energy dependence of the Brody parameter, which characterizes the
energy level statistics in the quantum system, is consistent with that of
the novel kind of MLE.
\end{abstract}

\pacs{05.45.Pq, 45.50.Jf, 73.21.Hb, 73.23.-b}
%Numerical simulations of chaotic systems

\maketitle

\section{Introduction}
The recent development in high technology has fabricated nano-scale 
quantum dots containing
a finite number of interacting electrons and optically trapped atoms 
where a finite number 
of interacting macroscopic particles are trapped in a small area. 
Quantum mechanics of 
these systems constitutes a topical subject.
In such systems, the underlying classical motion is expected to play an 
important role. 
The nature of the classical motion, i.e. regular, mixed, or
chaotic character, reflects on some of the quantum properties of
the systems, particularly in the energy level statistics. 
In this context, a number of studies on the quantum chaos of 
systems containing a few electrons
in quantum dots have been reported \cite{Ull,Mez,Ahn,Ahn2,Ves,
Fen,Dro,Xav,Saw}. 
However, closer correspondence between the classical chaos and the 
quantum chaos in 
those systems has not been investigated well. 

The simplest system among them would be the one-dimensional system 
\cite{Ves,Fen}.
In this paper we are concerned with behavior of two particles 
interacting with each other via the repulsive or attractive 
Coulomb potential in a one-dimensional 
system and study the correspondence between the classical and quantum 
chaos in detail. 
According to Fendrik et. al. \cite{Fen}, we introduce an effective Hamiltonian 
for a quantum wire, 
which reduces the original 3D system to the quasi-one-dimensional 
system. 
While its classical dynamics has some scaling property, the quantum 
counterpart has 
no such scaling property. We, however, show that the energy level 
statistics 
implies the existence of the corresponding scaling property even in the 
quantum system. 
This is demonstrated by calculations of the Brody parameter for 
distributions of the 
nearest neighbor level spacing (NNLS). 
This subject, scaling in quantum chaos, has been examined for some other 
systems, 
coupled harmonic or quartic oscillators \cite{Hal,Zim,Zen} and the hydrogen 
atom in a magnetic 
field \cite{Win}.

In order to clarify closer correspondence between the classical and 
quantum chaos, 
we introduce a novel kind of maximum Lyapunov exponent (MLE) instead of 
ordinary MLE. The ordinary MLE is a measure of the rate per unit of time 
for separation 
between two adjacent orbits while the new MLE is the one per unit of 
distance for 
separation between them. 
The new MLE is a suitable measure to compare chaotic irregularity 
among classical orbits with different energies.
We show that tendency of the energy dependence of the Brody parameter is 
consistent 
with that of the new MLE. 
We further show that the area of a chaotic region in Poincar\'e maps 
are not a suitable measure of chaotic irregularity for the present 
system, 
while several authors showed that it is a suitable measure of the 
irregularity 
in other systems \cite{Win,Ter,Har}. 

This paper is organized as follows:
In Sec.\ref{Model and method}, we construct
a quasi-one-dimensional model of two electrons confined 
in a quantum wire. We introduce a new kind of MLE.
In Sec.\ref{Numerical results}, we explore the distribution of NNLS
in wide range of energy and interaction strength. Then we examine 
the chaotic irregularity of the corresponding classical system with 
the use of the MLE and Poincar\'e maps.
We clarify correspondence between the energy dependence of the 
distribution of NNLS and the chaotic irregularity in the classical counterpart.
Summary and conclusion are given in Sec.\ref{conclusion}

\section{Model and method}
\label{Model and method}
\subsection{Quantum dynamics}
\begin{figure}[tb]
  \begin{center}
    \includegraphics[width=7cm]{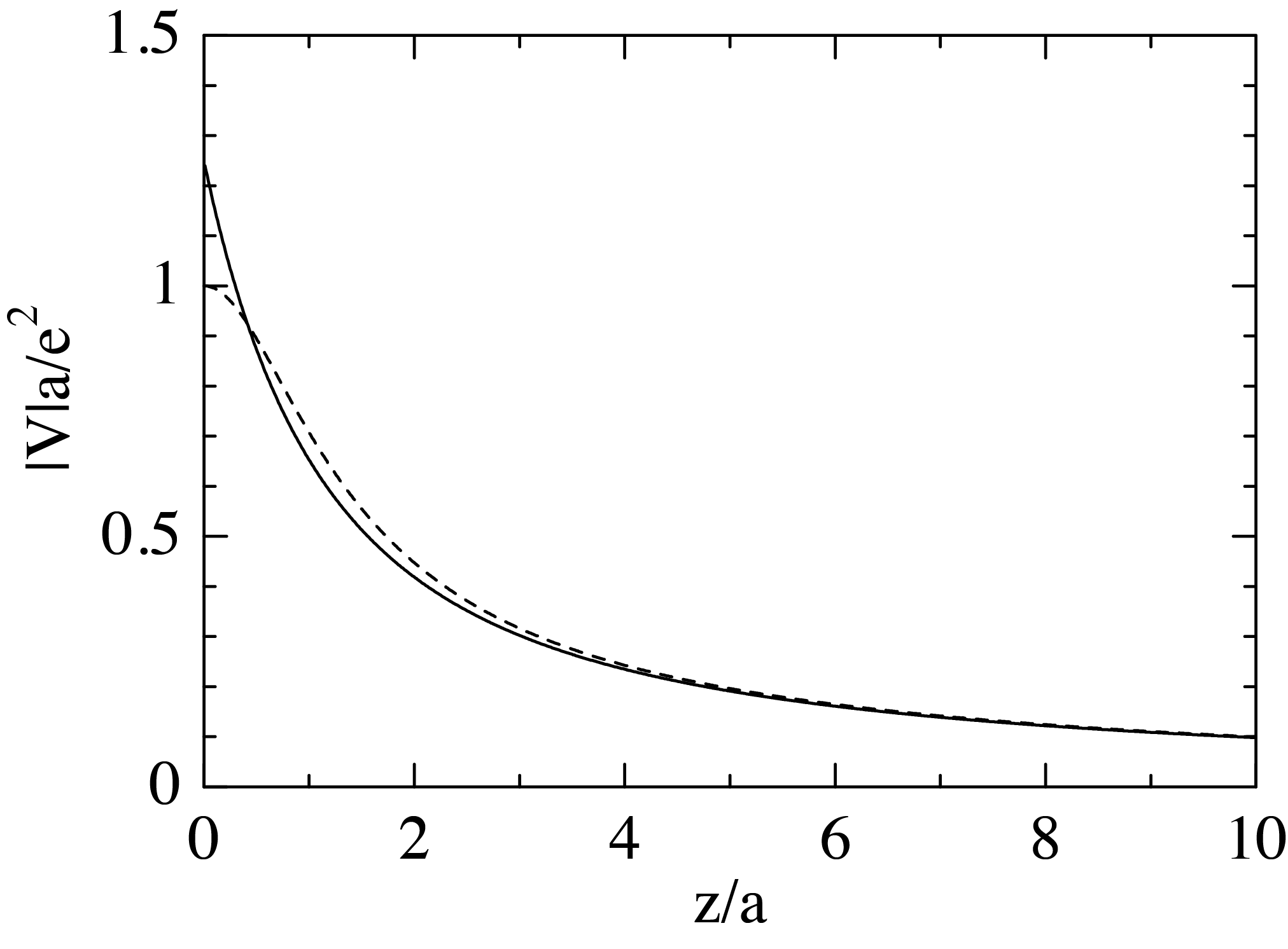}
  \end{center}
  \caption{The numerically calculated potential $V_{1D}$ (solid curve) 
and the analytical potential $V_{m}$ (broken curve). The lengths and 
potentials are scaled by $a$ and $e^2/a$, respectively.
}
\label{poten}
\end{figure}

We consider two spinless particles (two electrons or an electron-hole pair 
with the same mass) confined in a quantum wire. We assume a narrow parabolic 
confinement in the transversal directions ($x$ and $y$-directions), which are 
much narrower than a confinement in the longitudinal direction ($z$-direction).
We consider a hard wall potential in z-direction. 
The particles are interacting with each other via the repulsive or 
attractive Coulomb potential. 
The Hamiltonian of the system is written as
\begin{eqnarray}
%\begin{align}
H&=&\sum_{i=1,2}\Big{[}-\frac{\hbar^{2}}{2m}\left(\frac{\partial^{2}}
{\partial x_{i}^{2}}+\frac{\partial^{2}}{\partial y_{i}^{2}}+
\frac{\partial^{2}}{\partial z_{i}^{2}}\right)\nonumber\\
&&+\frac{1}{2}m\omega^{2}
(x_{i}^{2}+y_{i}^{2})\Big{]} \nonumber\\
&&\pm\frac{e^{2}}{\sqrt{(x_{1}-x_{2})^{2}+
(y_{1}-y_{2})^{2}+(z_{1}-z_{2})^{2}}}.\nonumber\\
\label{h1}
%\end{align}
\end{eqnarray}
We assume that the particles occupy the lowest-energy state associated 
with the transverse motion, which is energetically well separated from 
the excited states. Then the two-particle wave function can be 
approximated as
\begin{eqnarray}
\Psi(\bm{r}_{1},\bm{r}_{2})=\phi_{0}(x_{1})\phi_{0}(y_{1})\phi_{0}
(x_{2})\phi_{0}(y_{2})\Phi(z_{1},z_{2}),\nonumber\\
\label{h2}
\end{eqnarray}
where $\phi_{0}(x)$ is the lowest energy eigenstate of a 
harmonic oscillator. 
The wave function $\Phi(z_{1},z_{2})$ satisfies the equation
\begin{eqnarray}
H_{1D}\Phi(z_{1},z_{2})=E\Phi(z_{1},z_{2}),
\end{eqnarray}
where the effective Hamiltonian $H_{1D}$ is defined by
\begin{eqnarray}
H_{1D}=-\frac{\hbar^{2}}{2m}\left(\frac{\partial^{2}}
{\partial z_{1}^{2}}+\frac{\partial^{2}}{\partial z_{2}^{2}}\right)
+V_{1D}(\left|z_{1}-z_{2}\right|).\nonumber\\
\label{h3}
\end{eqnarray}
$V_{1D}(z)$ is the effective potential given by
%\begin{eqnarray}
\begin{widetext}
\begin{align}
V_{1D}(z)&=\pm e^{2}\int\frac{|\phi_{0}(x_{1})|^2|\phi_{0}
(y_{1})|^2|\phi_{0}(x_{2})|^2|\phi_{0}(y_{2})|^2}{\sqrt{(x_{1}-x_{2})^{2}
+(y_{1}-y_{2})^{2}+z^{2}}}dx_{1}dy_{1}dx_{2}dy_{2} \nonumber\\
&=\pm\frac{e^{2}}{a}\int_{0}^{\infty}\frac{s\exp[-s^{2}/2]}{\sqrt{s^{2}
+(z/a)^{2}}}ds
\end{align}
\end{widetext}
%\end{eqnarray}
and $a=\sqrt{\hbar/m\omega}$. 
In this way, our system exhibits a quasi-one-dimensional property. 
Now we introduce a model potential \cite{Fen} defined by
\begin{eqnarray}
V_{m}(z)=\pm\frac{e^{2}}{\sqrt{a^{2}+z^{2}}}.
\end{eqnarray}
In Fig.\ref{poten}, the solid and broken curves indicate the numerically 
calculated potential $V_{1D}$ and the analytical potential 
$V_{m}$, respectively. 
We see that $V_{1D}$ can be well approximated by $V_{m}$. 
We adopt $V_{m}$ instead of $V_{1D}$ as the interaction 
potential between particles since the analytical potential can be 
dealt with more easily. 
Thus our effective Hamiltonian is written as
\begin{eqnarray}
H_{eff}&=&-\frac{\hbar^{2}}{2m}\Big{(}\frac{\partial^{2}}{\partial z_{1}^{2}}
+\frac{\partial^{2}}{\partial z_{2}^{2}}\Big{)}\nonumber\\
&&\pm\frac{e^{2}}{\sqrt{a^{2}
+(z_{1}-z_{2})^{2}}}.
\label{h4}
\end{eqnarray}
%%%%%%%%
%%%%%%%%

We scale lengths, angular momentums and masses by $L$, $\hbar$ and $2m$,
respectively, 
where $L$ is a width of the system in the longitudinal direction. 
Then the effective Hamiltonian is reduced to
\begin{eqnarray}
H_{eff}=-\left(\frac{\partial^{2}}{\partial z_{1}^{2}}+\frac{\partial^{2}}
{\partial z_{2}^{2}}\right)+\frac{\lambda}{\sqrt{(z_{1}-z_{2})^{2}
+\delta^{2}}},\nonumber\\
\label{h5}
\end{eqnarray}
where $\lambda$ is the effective interaction strength parameter given by 
$\lambda=\pm 2e^{2}mL/\hbar^{2}$ and $\delta=a/L$. 
Note that the parameter $\lambda$ can be expressed as $\lambda=\pm 2L/a_{B}$, 
where $a_{B}$ is the Bohr radius defined by $a_{B}=\hbar^{2}/e^{2}m$.
The particles are confined by hard walls within $0\leq z_{1}\leq 1$ and 
$0\leq z_{2}\leq 1$. These hard walls describe the boundary of the 
quasi-one-dimensional wire. 
We examine the feature of the system as varying $\lambda$ while keeping 
$\delta$, which implies that we vary the system size keeping the ratio 
between the longitudinal and transversal lengths.

It should be noted that the present 1D two-particle system is equivalent 
to a 2D system of one particle having a coordinate $(z_{1}, z_{2})$ 
within a hard-walled square billiard. 
The Hamiltonian of the latter system is also given by Eq.(\ref{h5}), 
in which the first and second terms represent the kinetic energy of 
the particle and the third term represents a external potential. 
We can chose energy eigenfunctions of the 2D system as being symmetric or 
antisymmetric against exchange between $z_{1}$ and $z_{2}$, i.e.,
$\Phi(z_{1},z_{2})=\Phi(z_{2},z_{1})$ or $\Phi(z_{1},z_{2})=
-\Phi(z_{2},z_{1}),$ since the Hamiltonian does not change under this exchange. 
The symmetric and antisymmetric cases correspond to the boson and fermion 
cases, respectively, in the 1D two-particle system. 
In the present paper we  are concerned only with the cases of fermions.

In order to look for signatures of quantum chaos in the present system, 
we examine distributions of the nearest neighbor level spacing (NNLS). 
The eigenenergies are obtained by diagonalizing the Hamiltonian matrices 
numerically, whose elements are evaluated by using energy eigenstates 
without the Coulomb interaction (Slater determinants) as a basis set:
\begin{eqnarray}
\phi_{m,n}(z_1,z_2) &=& \sqrt{2}
\Big{(} \sin(m\pi z_1)\sin(n\pi z_2) \nonumber\\
&&- \sin(n\pi z_1)\sin(m\pi z_2) \Big{)},
\label{phimn1}
\end{eqnarray}
where $m$ and $n$ are integer larger than zero. 
The components of the Hamiltonian matrices are represented
with respect to $\phi_{m,n}$ in Eq.(\ref{phimn1}) as
\begin{eqnarray}
&&<\phi_{m,n}|H|\phi_{m',n'}>
=\pi^2(m^2+n^2)\delta_{m,m'}\delta_{n,n'}\nonumber\\
&&
\hspace{0.2cm}+4\lam\{I(m,n|m',n')-I(m,n|n',m')\},
\end{eqnarray}
where $I(m,n|m',n')$ is defined by
\begin{widetext}
\begin{eqnarray}
I(m,n|m',n') = \int_0^1 \int_0^1dz_1dz_2
\frac{\sin(m\pi z_1)\sin(n\pi z_2)
\sin(m'\pi z_1)\sin(n'\pi z_2)}{\sqrt{(z_1-z_2)^2+\delta^2}}.\nonumber\\
\end{eqnarray}
\end{widetext}
We further take into account the parity of the system. 
The present system is invariant under the inversion associated with the 
center $(z_{1},z_{2})=(1/2,1/2)$. 
Therefore, the eigenstates are classified into ones having the even parity 
with $(m,n)=$(even, even) or (odd, odd) and those having odd parity with 
$(m,n)=$(even, odd) or (odd, even). We concentrate ourselves on the 
eigenstates of the even parity in this paper, when we examine NNLS.

NNLS is fitted to the Brody distribution function
\begin{eqnarray}
P_B(S)&=& (\alpha+1)b S^\alpha \exp(-bS^{\alpha+1}),\nonumber\\
b &=& \Big\{\Gamma\Big(\frac{\alpha+2}{\alpha+1}\Big) \Big\}^{\alpha+1},
\label{bro}
\end{eqnarray}
which interpolates the Poisson and Wigner distributions. 
It coincides with the Poisson distribution for $\alpha=0$ and recovers 
the Wigner distribution for $\alpha=1$. We use the Brody parameter 
$\alpha$ as a measure for degree of chaotic irregularity of the system.

\subsection{Classical dynamics}
Now we turn to the dynamics of the classical counterpart of the two-particle 
system described by the Hamiltonian (\ref{h5}) .
Similarly to the quantum case, lengths, angular momentums and masses
are scaled by
$L$, $\hbar$ and $2m$, respectively.
%We also scale lengths, energies and angular momentums by $L,\ \hbar^{2}/2mL^{2}$ 
%and $\hbar$, respectively. (Note that masses are scaled by $2m$.) 
%\textcolor{blue}{The scaling is consistent with the scaling of length and energy for the 
%Quantum system in the previous subsection.　前の量子力学でスケールした系と同じだということを述べた方が良いかと思い
%書き足してみましたがいかがでしょう。consistent という単語が適切かわかりません。}
The equations of the motion are then given as
\begin{eqnarray}
\frac{1}{2}\frac{d^2z_1}{dt^2}
=\frac{\lam(z_1-z_2)}{\{(z_1-z_2)^2+\delta^2\}^{3/2}},\nonumber\\
\frac{1}{2}\frac{d^2z_2}{dt^2}
=\frac{\lam(z_2-z_1)}{\{(z_1-z_2)^2+\delta^2\}^{3/2}}.
\label{h1_2}
\end{eqnarray}
The $\lam$ defined by $\lambda=\pm 2e^{2}mL/\hbar^{2}$ is dimensionless.
The total energy is represented as 
\begin{eqnarray}
E&=&\frac{1}{4}\Big{(}\frac{dz_1}{dt}\Big{)}^2 + 
\frac{1}{4}\Big{(}\frac{dz_2}{dt}\Big{)}^2\nonumber\\
&&+\frac{\lam}{\sqrt{(z_1-z_2)^2+\delta^2}}.
\end{eqnarray}
The dynamics of the present system is equivalent to that of a particle 
confined in a two-dimensional square box in $0\leq z_{1}\leq 1$ and 
$0\leq z_{2}\leq 1$ with the potential $\lambda/\sqrt{(z_{1}-z_{2})^{2}
+\delta^{2}}$ similarly to the quantum case. 
The behavior of the system apparently depends on the interaction 
strength parameter $\lambda$. 
However, if we introduce a rescaled time $\tau$ defined by 
$\tau=\sqrt{|\lambda|}t$, the equations of the motion are reduced to
\begin{eqnarray}
\frac{1}{2}\frac{d^2z_1}{d\tau^2}
=\pm\frac{(z_1-z_2)}{\{(z_1-z_2)^2+\delta^2\}^{3/2}},\nonumber\\
\frac{1}{2}\frac{d^2z_2}{d\tau^2}
=\pm\frac{(z_2-z_1)}{\{(z_1-z_2)^2+\delta^2\}^{3/2}}.
\label{eq7_7_1}
\end{eqnarray}
The rescaled energy $\ep=E/|\lambda|$ is given by 
\begin{eqnarray}
\ep&=&\frac{1}{4}\Big{(}\frac{dz_1}{d\tau}\Big{)}^2 + 
\frac{1}{4}\Big{(}\frac{dz_2}{d\tau}\Big{)}^2\nonumber\\
&&\pm\frac{1}{\sqrt{(z_1-z_2)^2+\delta^2}}.
\label{eq7_7_2}
\end{eqnarray}
The plus and minus signs in Eqs.(\ref{eq7_7_1}) and (\ref{eq7_7_2})
correspond to positive and negative $\lam$, respectively. 
Consequently, the classical behavior of the system is independent of 
the value of $\lambda$ itself. This is because in the classical system 
there is no such characteristic length as the Bohr radius due to the 
finite Plank constant in the quantum system. 
Even if we enlarge the system size, we can find the equivalent trajectory 
by increasing the total energy. 
On the other hand, the quantum behavior of the system depends on 
the value of $\lambda$ directly. Now the following question arises: 
Even in the quantum case, whether does the system 
with the same value of $E/|\lambda|$ but with different $\lambda$
exhibit a similar behavior to classical system, 
especially, concerning the degree of chaotic 
irregularity of the system?
Otherwise there is no closer 
correspondence between the classical chaos and the quantum chaos. 
We investigate this point in the present study. 

%MLE is one of a measure of 
%the degree of sensitivity to initial conditions. 
%MLE is rate of exponential convergence or divergence of 
%nearby orbits of a dynamical system in phase space.
%Positive MLE shows the divergence of nearby trajectories
%representing the sensitivity of the system on initial conditions. 
%The maximum MLE distinguishes periodic behavior from chaos.

\begin{figure}[h!]
\begin{center}
\includegraphics[width=5cm]{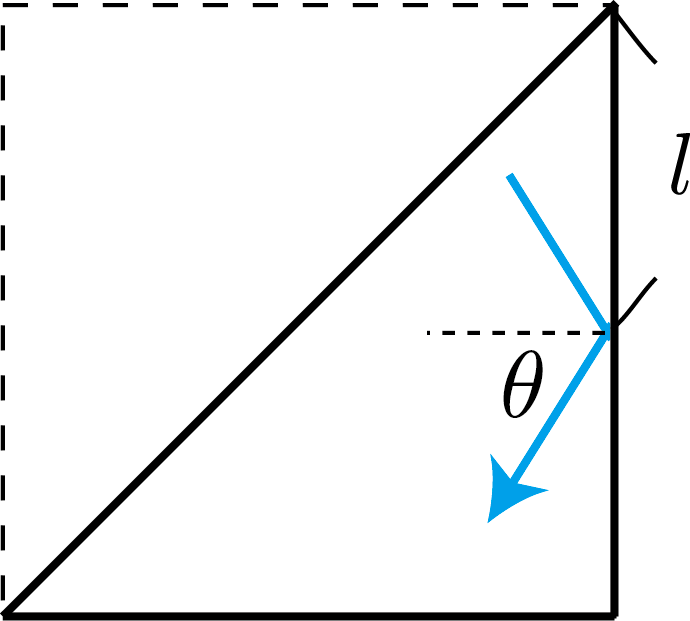}% Here is how to import EPS art
\end{center}
\caption{\label{fig:epsart}The coordinates of the billiard system.
}
\label{poin}
\end{figure}

We use two kinds of the Poincar\'e map to see behavior of the classical
system.
The first kind of Poincar\'e map called Poincar\'e map 1
is defined in the section $v_2$ versus
$z_2$ for the second particle taken at times when the first particle bounces
off the left boundary of the well $(z_1=0)$.
The second kind of Poincar\'e map called Poincar\'e map 2 
is defined as follows.
We take coordinate $l$ along the
two sides of the square and a diagonal line connecting $(z_1,z_2)=(0,0)$ and 
$(1,1)$ where the ridge of the potential lies (see Fig.\ref{poin}). 
$l$ is normalized so that the range is $0\le l \le 1$.
The trajectory can be recorded by two values.
One is $l$ at point where the particle is reflected on the hardwalls 
or intersects the 
line $z_1=z_2$.
The other is $p=\cos\theta$, where $\theta$ is the angle 
between the velocity vector after reflection and the normal to the
solid line. %in Fig.{\ref{poin}}.
The Poincar\'{e} map which records $l$ and $\theta$ of orbits
reflects the property of the classical system.

We also use MLE as a measure 
for degree of chaotic irregularity
of the classical system.
The ordinary Lyapunov exponent is defined as follows:
we consider an orbit $(\largep(\tau),\largeq(\tau))$ 
(denoted as the reference orbit)
and a slightly displaced orbit from the reference orbit
in the phase space.
The starting point of the displaced orbit is spaced apart by a small vector
$(\Delta\largep(0),\Delta\largeq(0))$ from $(\largep(0),\largeq(0))$ 
at initial time $\tau=0$.
The distance between the reference and displaced orbits is
\begin{eqnarray}
d_0=|(\Delta \largep(0),\Delta\largeq(0))|.
\label{lia1}
\end{eqnarray}
We follow these orbits for a time interval $\Delta \tau$.
The distance between the two orbits at $\tau=\Delta \tau$ is represented as
\begin{eqnarray}
d_1=|(\Delta \largep(\Delta \tau),\Delta\largeq(\Delta \tau))|.
\label{lia2}
\end{eqnarray}
Then we choose a new starting point of displaced trajectory at time 
$\tau=\Delta \tau$ as
\begin{eqnarray}
&&(\largep(\Delta \tau),\largeq(\Delta \tau)) \nonumber\\
&&\hspace{0.5cm}+ \frac{d_0}{d_1}
(\Delta\largep(\Delta \tau),\Delta\largeq(\Delta \tau))
\end{eqnarray}
so that the distance between the new starting points 
equals $d_0$.
The trajectory is followed up to time $\tau=2\Delta \tau$.
The new deviation of the displaced orbit from the reference orbit  
\begin{eqnarray}
d_2=|(\Delta \largep(2\Delta \tau),\Delta\largeq(2\Delta \tau))|
\label{lia3}
\end{eqnarray}
is computed, and a second rescaled trajectory is started.
This process is continued, yielding a sequence of distances
$d_0,d_1,d_2,\cdots$.
By using these values, MLE is defined as
\begin{eqnarray}
\Gamma_L = \lim_{n\rightarrow\infty}\frac{1}{n\Delta \tau}
\sum_{i=1}^{n}\ln\frac{d_i}{d_0},
\label{lia4}
\end{eqnarray}
where $n$ is the number of the time segment.
%Unfortunately the MLE in Eq.(\ref{lia4}) 
%defined for the equations of motion (\ref{h1_2})
%does not 
%reflect the property that 
%the classical system is characterized by the one parameter $\ep=E/|\lam|$.
%$\Gamma_L$ can take various values depending on $E$ for fixed $\ep$.
%One candidate of the MLE independent $E$ or $\lam$ themselves 
%is the MLE defined for the equation of motion (\ref{eq7_7_1})
%which depends on $\ep$ but not on $E$ or $\lam$ themselves.
This quantity $\Gamma_L$ is, however, 
not suitable as a measure for degree of chaotic 
irregularity in the present system.
$\Gamma_L$
increases with $\ep$ only even 
because the motion of particles becomes faster
with the increase of $\ep$, 
while the classical dynamics becomes regular 
in high energy regime as shown by the numerical results 
in the next section. 

We introduce a novel kind of 
MLE defined as
\begin{eqnarray}
\Gamma'_L 
= \lim_{n\rightarrow\infty}\sum_{i=1}^{n}\frac{1}{n\xi}\ln\frac{d_i}{d_0}.
\label{lia5}
\end{eqnarray}
The definition of $\Gamma_L'$ is almost the same as $\Gamma_L$
except that $\Delta\tau$ in Eq.(\ref{lia4}) is replaced by a small 
distance $\xi$.
The definition of $\Gamma'_L$ in Eq.(\ref{lia5}) is similar to Eq.(\ref{lia4}).
%To evaluate $\Gamma_L$ in Eq.(\ref{lia4}) 
%we follow two adjacent orbits (reference and displaced orbits)
%during the time interval $\Delta\tau$ and rescale the displaced orbit
%so that the distance between the new starting points 
%equals $d_0$.
For $\Gamma_L'$ 
we follow the two adjacent orbits while the reference orbit
travels small distance $\xi$, and then evaluate $d_i$.
$\Gamma'_L$ represents the degree of exponential divergence of adjacent orbits
similarly to $\Gamma_L$.
However, $\Gamma'_L$ depends only on geometry of orbits but not 
on quickness of development of orbits.
We employ $\Gamma'_L$ in Eq.(\ref{lia5}) 
as a measure of degree of chaotic irregularity in classical mechanics.

\section{Numerical results}
\label{Numerical results}
\subsection{Repulsive interaction}
\begin{figure}[h!]
\begin{center}
\includegraphics[width=7cm]{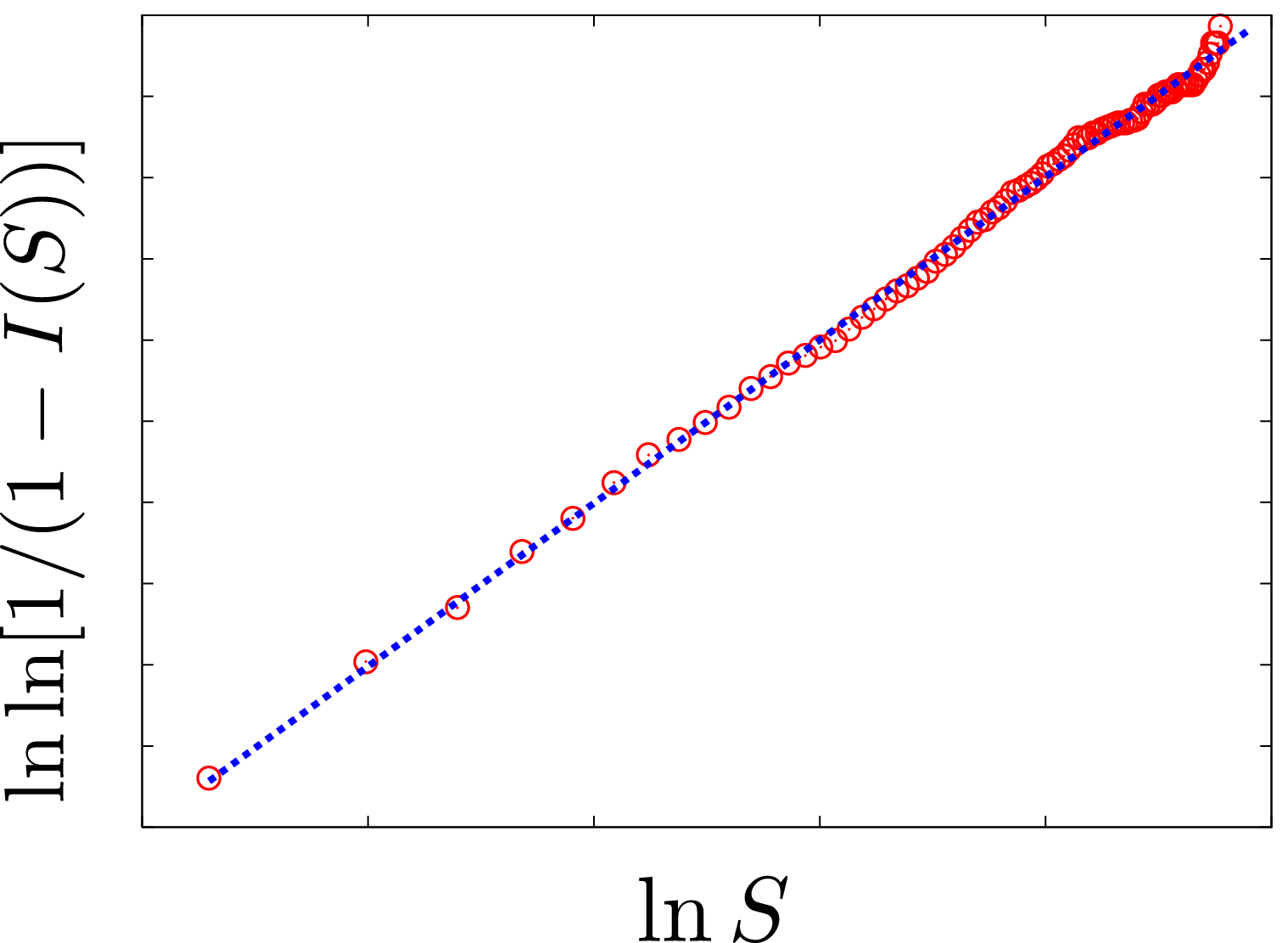}% Here is how to import EPS art
\end{center}
\caption{\label{fig:epsart}Relation between $\ln S$ and $\ln\ln[1/(1-I(S))]$
for $\lam=200$ and $\delta=0.01$. 
$I(S)$ is the cumulative number of states in the unfolded spectrum.
%an integral defined in Eq.(\ref{IB}) with $P(S)$ instead of $P_B(S)$.
The NNLS is calculated by using the $1000$th eigenstate and following 
$1000$ eigenstates.
The dotted line is obtained by using Eq.(\ref{eq_8_4_1}) and
the least-squares fitting method .
}
\label{fit_lam_200_1000_2000}
\end{figure}

We obtain energy eigenvalues by diagonalizing the effective 
Hamiltonian (\ref{h5}), and 
evaluate the smoothed counting function $N_{av}(E)$
which gives the cumulative number of states below an energy $E$.
In order to analyze the structure of the level-fluctuation properties,
we unfold the spectrum by applying the well-known transformation $x_n
= N_{av}(E_n)$ to obtain a constant mean spacing, % \cite{Boh,Boh2},
where $n$ denotes the number of the energy level.
From the unfolded spectrum we obtain the histogram of 
the NNLS distribution $P(S)$, where $S_n=x_{n+1}-x_{n}$.
The histogram is fitted to the Brody distribution function $P_B(S)$ in 
Eq.(\ref{bro}).
The integral of the Brody distribution function, 
\begin{eqnarray}
I_B(S)\equiv \int_0^SP_B(S')dS'
\label{IB}
\end{eqnarray}
satisfies 
\begin{eqnarray}
&&\ln\ln[1/(1-I_B(S))] \nonumber\\
&&\hspace{0.5cm}= (1+\alpha)\ln S + \ln b,
\label{eq_8_4_1}
\end{eqnarray}
where $b$ is given in Eq.(\ref{bro}).
By using the above relation and the least-squares fitting method 
we evaluate the Brody parameter $\alpha$ for the dstribution $P(S)$ of the
NNLS.
Hereafter we take $\delta=0.01$.
As an example, a result of fitting for 
$\lam=200$ is shown 
in Fig.\ref{fit_lam_200_1000_2000}.

\begin{figure}[tb]
  \begin{center}
    \includegraphics[width=8.0cm]{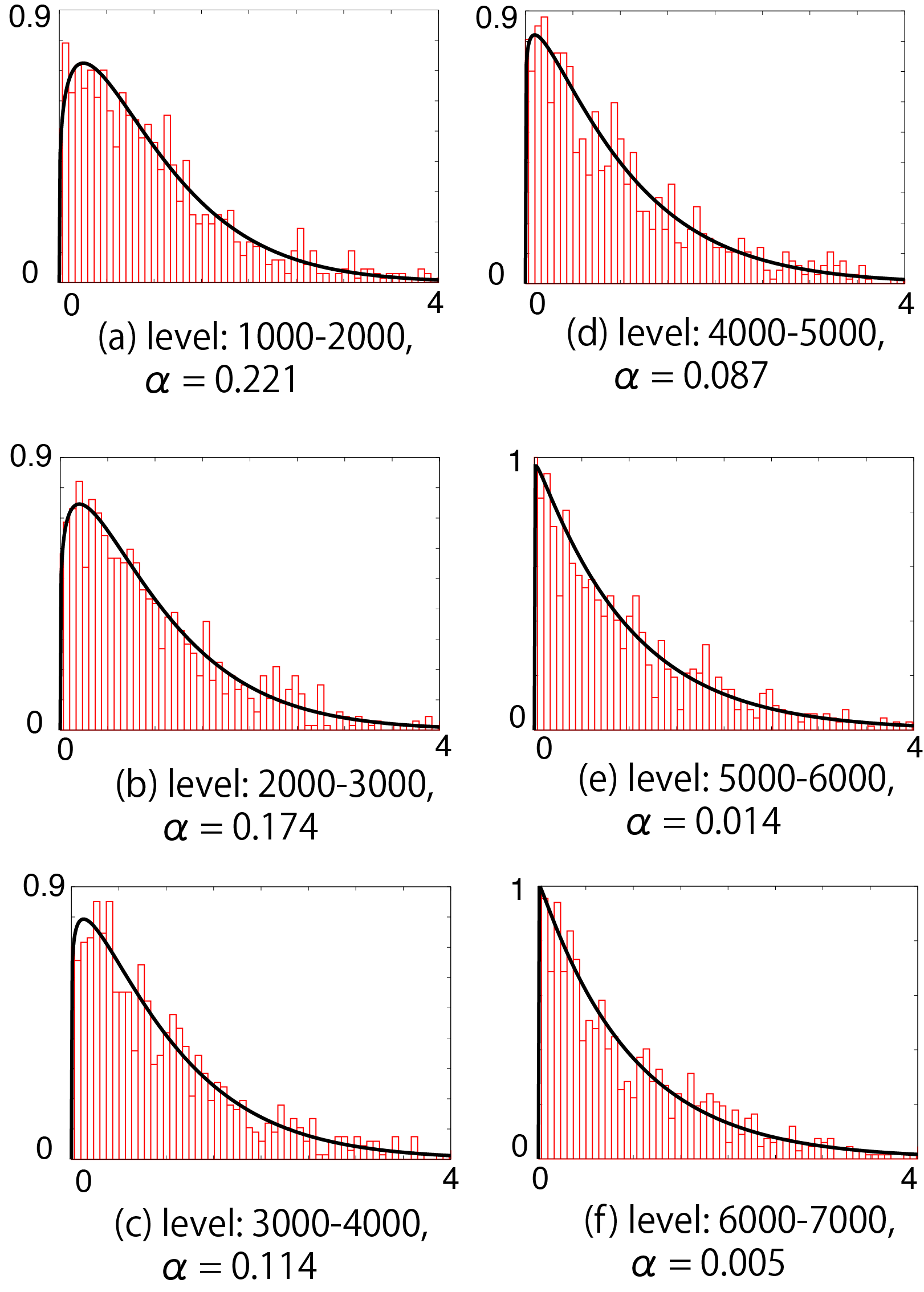}
  \end{center}
  \caption{Histograms of NNLS for $\lam=200$.
Solid lines are the best fitted Brody distributions.
Each value of the Brody parameter $\alpha$ is shown below each panel.
}
\label{dis_lam200}
\end{figure}

\begin{figure}[tb]
  \begin{center}
    \includegraphics[width=8.0cm]{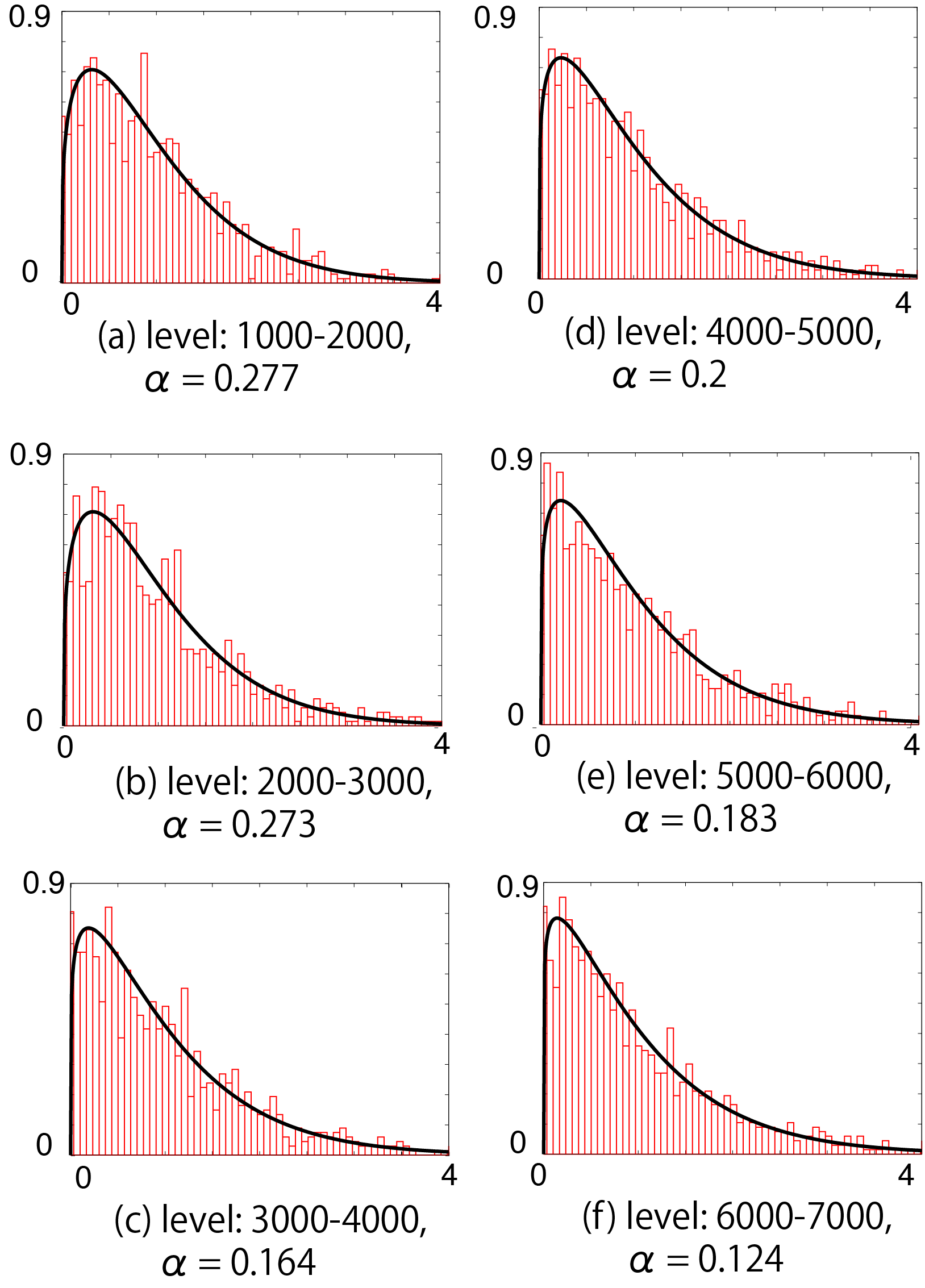}
  \end{center}
  \caption{
The same as Fig.\ref{dis_lam200} except for $\lam=500$.
}
\label{dis}
\end{figure}

The total energy region is divided into several regions.
In Fig.\ref{dis_lam200} we show the obtained NNLS
distribution in each region
for $\lam=200$.
About 1000 eigenvalues are used in each region to compute each histogram.
The range of the used energy levels and the 
Brody parameter $\alpha$ are shown below each panel.
We see that the Brody parameter decreases with increase of the average of
energy eigenvalues which are used to obtain the histogram.
Especially for the histograms in the panels (e) and (f) in Fig.\ref{dis_lam200}
with the Brody parameter less than $0.015$, the histograms are well fitted
also by the Poisson distribution.
The histograms for $\lam=500$ is shown in Fig.\ref{dis}.

We see that the Brody parameter $\alpha$ 
decreases with increase of the average of energy eigenvalues similarly to the case of $\lam=200$, and moreover that
the values of $\alpha$ are greater than those for $\lam=200$ 
in each energy level region.

\begin{figure}[h!]
\begin{center}
\includegraphics[width=8cm]{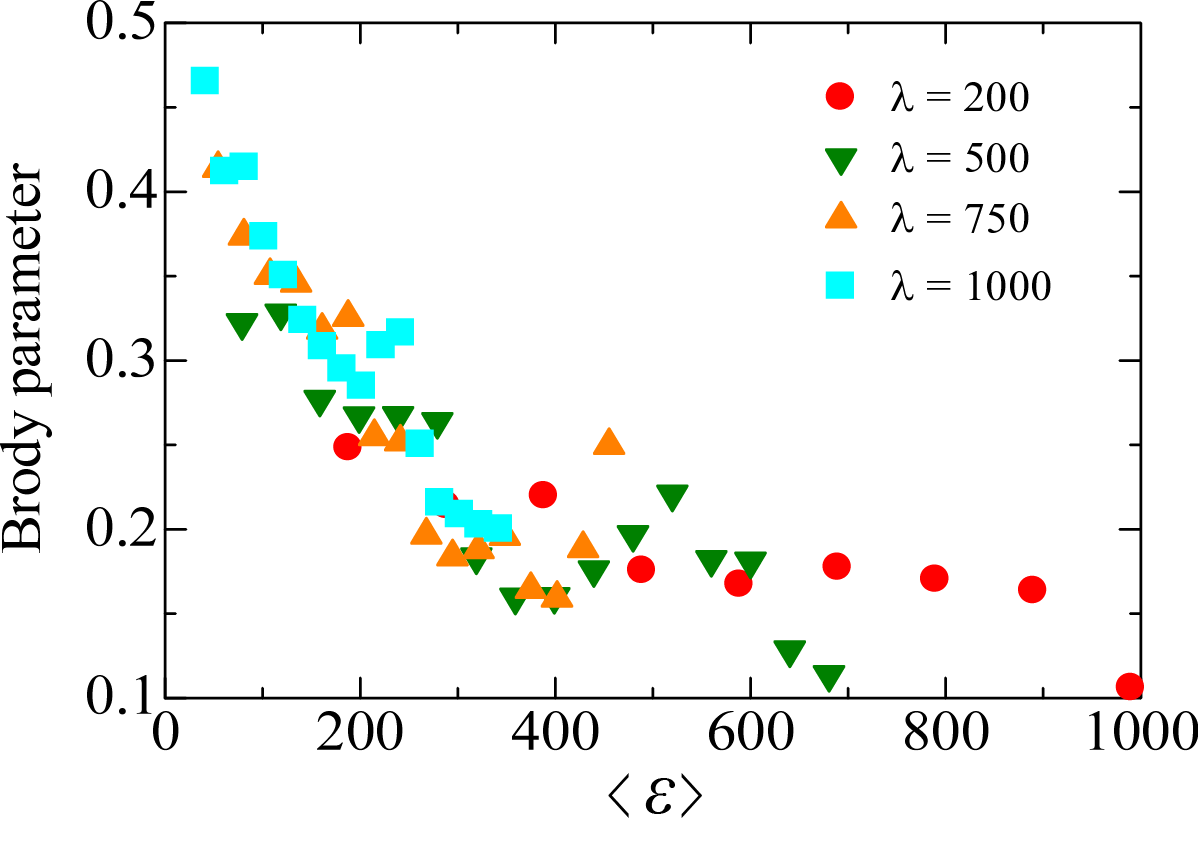}% Here is how to import EPS art
\end{center}
\caption{\label{fig:epsart}$\ep-$dependence of the Brody parameter for 
the repulsive interaction.
For each data point eigenvalues in the energy interval $\Delta E=50000$ 
were used (Their number is about 1000).
Horizontal axis denotes the average of $\ep$.
}
\label{bro_store_seki}
\end{figure}

We show the $\ep-$dependence of the Brody parameter for $\lam=200, 500,
750, 1000$ in Fig.{\ref{bro_store_seki}}.
For each data point we use energy eigenvalues in energy interval $\Delta E=
50000$ which includes about $1000$ energy eigenvalues.
The horizontal axis denotes the averaged value of $\ep$ of the 
used eigenstates, $<\ep>$.
We see that the Brody parameter $\alpha$ decreases almost monotonously with
increase of $<\ep>$.
Moreover the $\ep-$dependences of $\alpha$ is quite similar for different
$\lam$ especially in $<\ep> < 600$.
As mentioned in the previous section,
the classical system 
has scaling property characterized by parameter $\ep=E/|\lam|$.
The above results indicate that the distribution of 
NNLS in quantum mechanics has the same scaling property on $\ep$.

\begin{figure}[h!]
\begin{center}
\includegraphics[width=7cm]{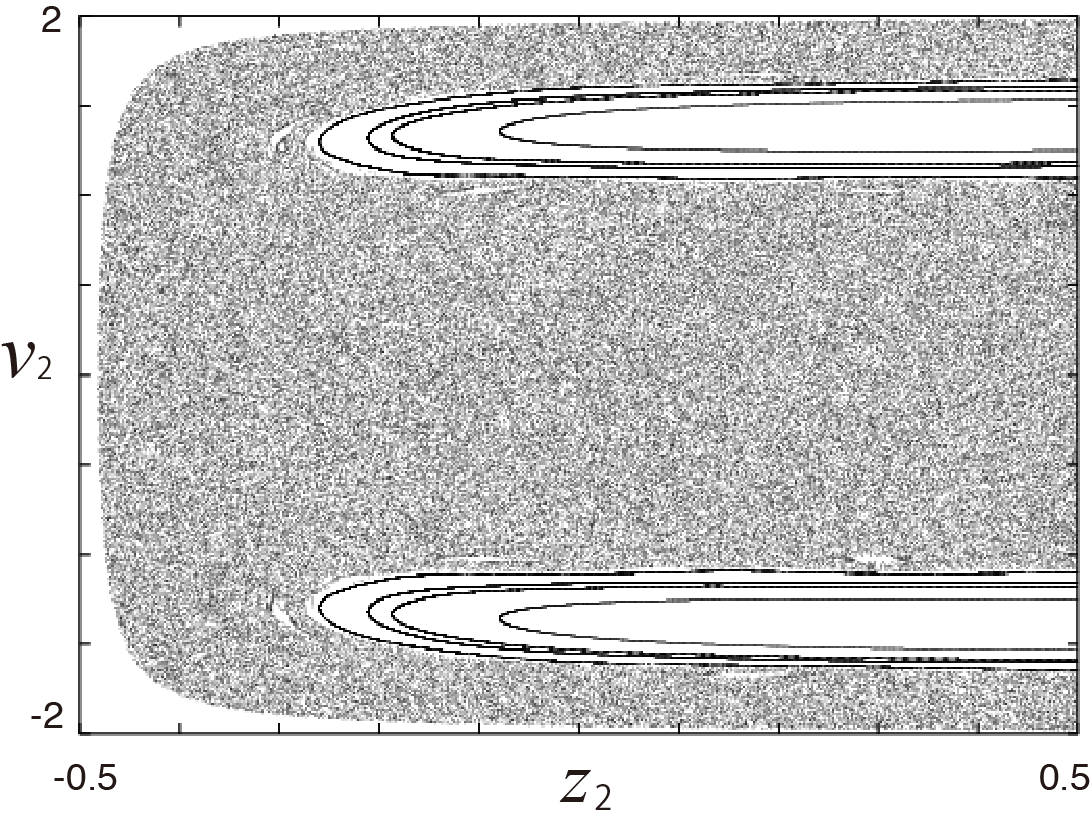}% Here is how to import EPS art
\end{center}
\caption{\label{fig:epsart}
Poincar\'e map 1 for repulsive interaction with $\ep=50$.
}
\label{map_E50_jpg}
\end{figure}

\begin{figure}[h!]
\begin{center}
\includegraphics[width=7cm]{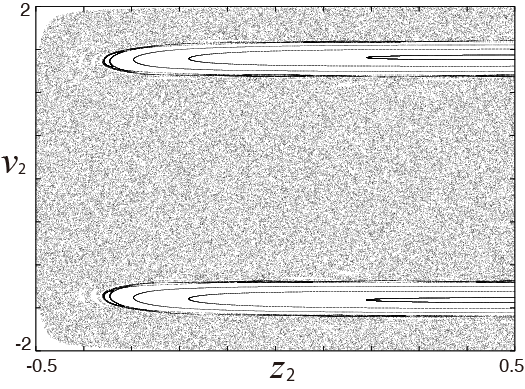}% Here is how to import EPS art
\end{center}
\caption{\label{fig:epsart}
Poincar\'e map 1 for repulsive interaction with $\ep=200$.
}
\label{map_E200_jpg}
\end{figure}

\begin{figure}[h!]
\begin{center}
\includegraphics[width=7cm]{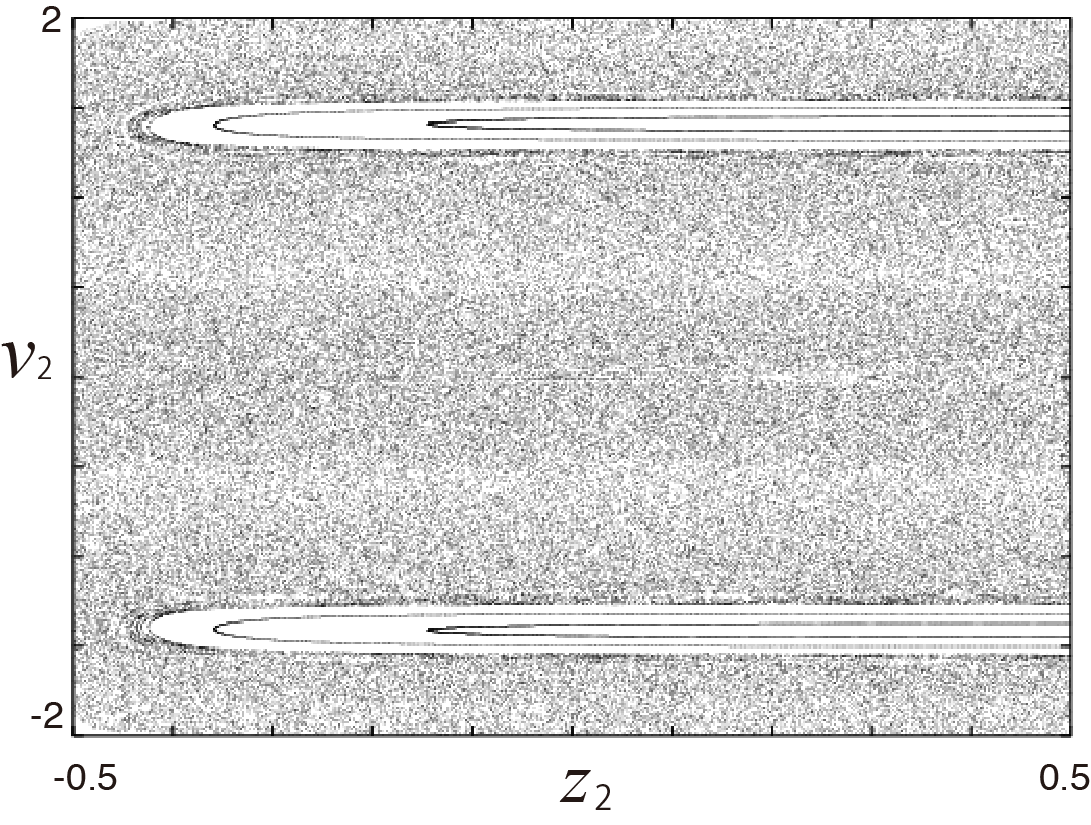}% Here is how to import EPS art
\end{center}
\caption{\label{fig:epsart}
Poincar\'e map 1 for repulsive interaction with $\ep=1000$.
}
\label{map_E1000_jpg}
\end{figure}

\begin{figure}[h!]
\begin{center}
\includegraphics[width=7cm]{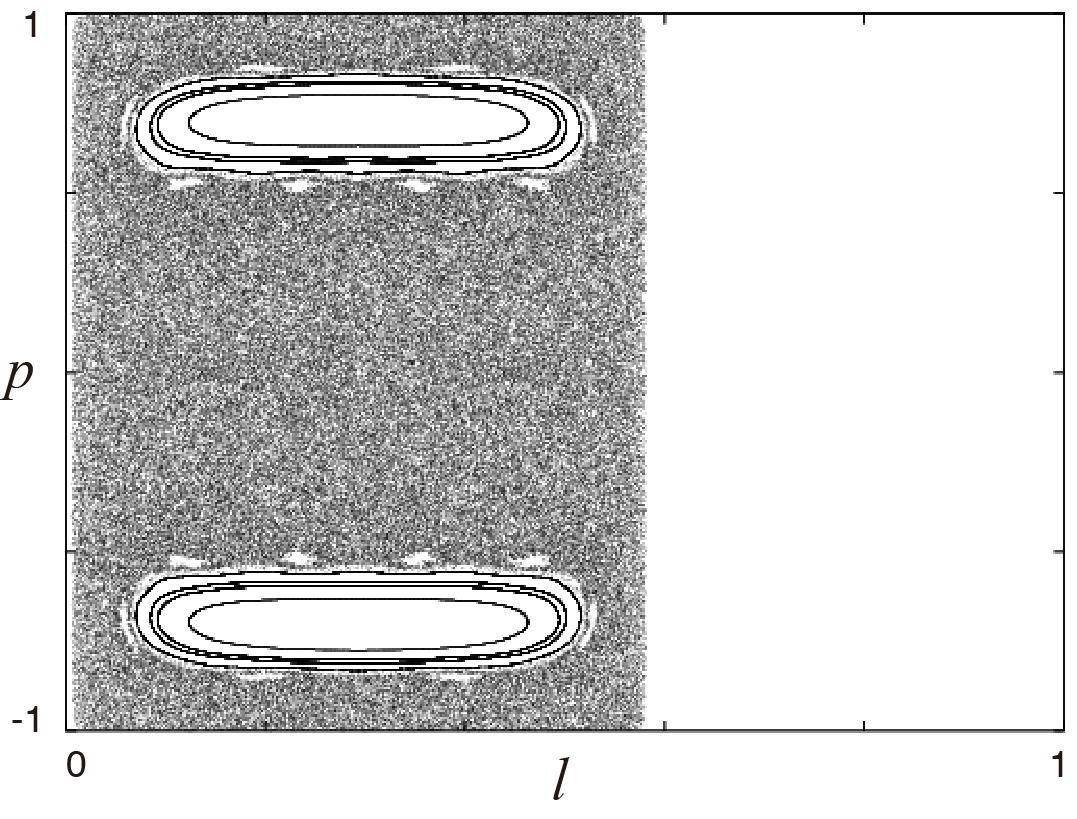}% Here is how to import EPS art
\end{center}
\caption{\label{fig:epsart}
Poincar\'e map 2 for repulsive interaction with $\ep=50$.
}
\label{map_E50_Bro}
\end{figure}

\begin{figure}[h!]
\begin{center}
\includegraphics[width=7cm]{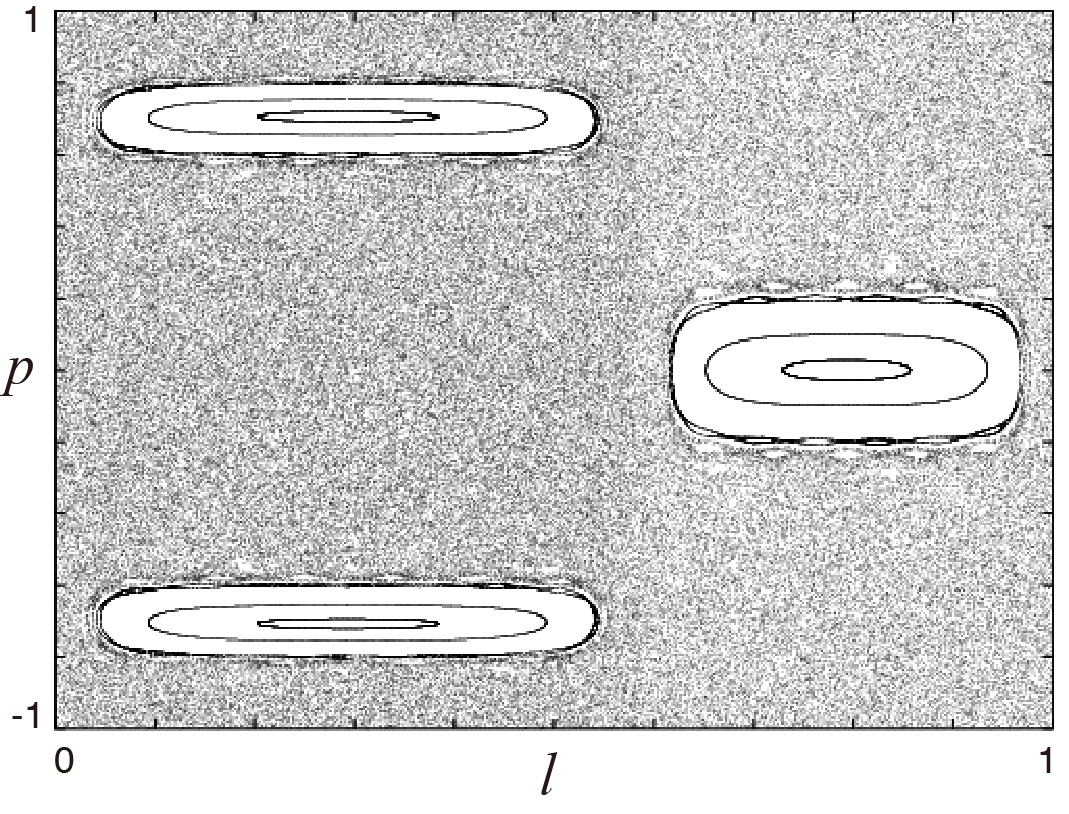}% Here is how to import EPS art
\end{center}
\caption{\label{fig:epsart}
Poincar\'e map 2 for repulsive interaction with $\ep=200$.
}
\label{map_E200_Bro}
\end{figure}

\begin{figure}[h!]
\begin{center}
\includegraphics[width=7cm]{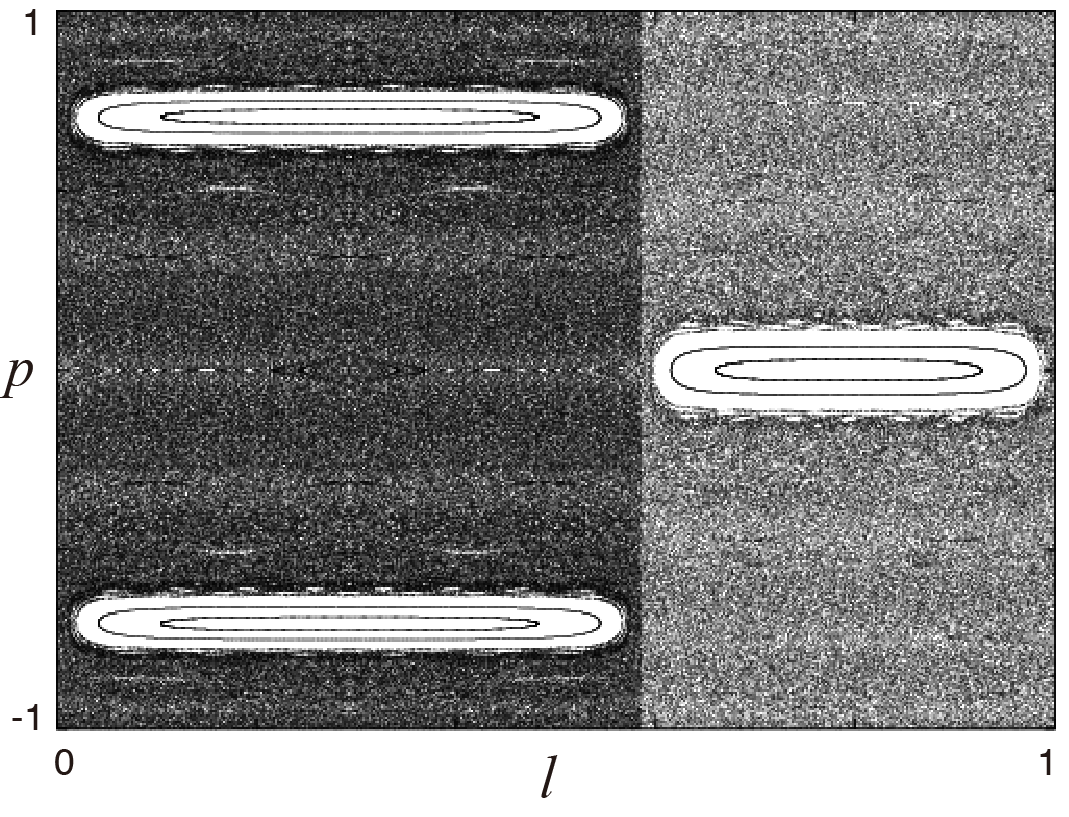}% Here is how to import EPS art
\end{center}
\caption{\label{fig:epsart}
Poincar\'e map 2 for repulsive interaction with $\ep=1000$.
}
\label{map_E1000_Bro}
\end{figure}

Now we see the behavior of the classical system
with the equations of motion,
Eq.(\ref{eq7_7_1}).
%In Figs.\ref{maps_lam75_2} -
%\ref{maps_lam75_2}, \ref{maps_lam100_2},
%\ref{maps_lam150_2}, \ref{maps_lam250_2}, \ref{maps_lam350_2}, \ref{maps_lam500_2},
%\ref{maps_lam700_2} 
In Figs.\ref{map_E50_jpg}-\ref{map_E1000_jpg} 
we show the Poincar\'e maps 1 defined in previous section
for $\ep= 50, 200, 1000$, respectively, with $\delta=0.01$.
In Figs.\ref{map_E50_Bro}-\ref{map_E1000_Bro}
the Poincar\'e maps 2 are shown.
We have taken about 20 different initial points in phase space for each map.
%Some orbits show the chaotic behavior in the Poincar\'{e} map.
These Poincar\'{e} maps show that the present classical system
exhibits mixed dynamics with coexisting KAM tri and chaotic regions.
This is consistent with the fact that NNLS distributions in the
quantum system are intermediate between the Poisson and Wigner
distribution.

\begin{figure}[h!]
\begin{center}
\includegraphics[width=6cm]{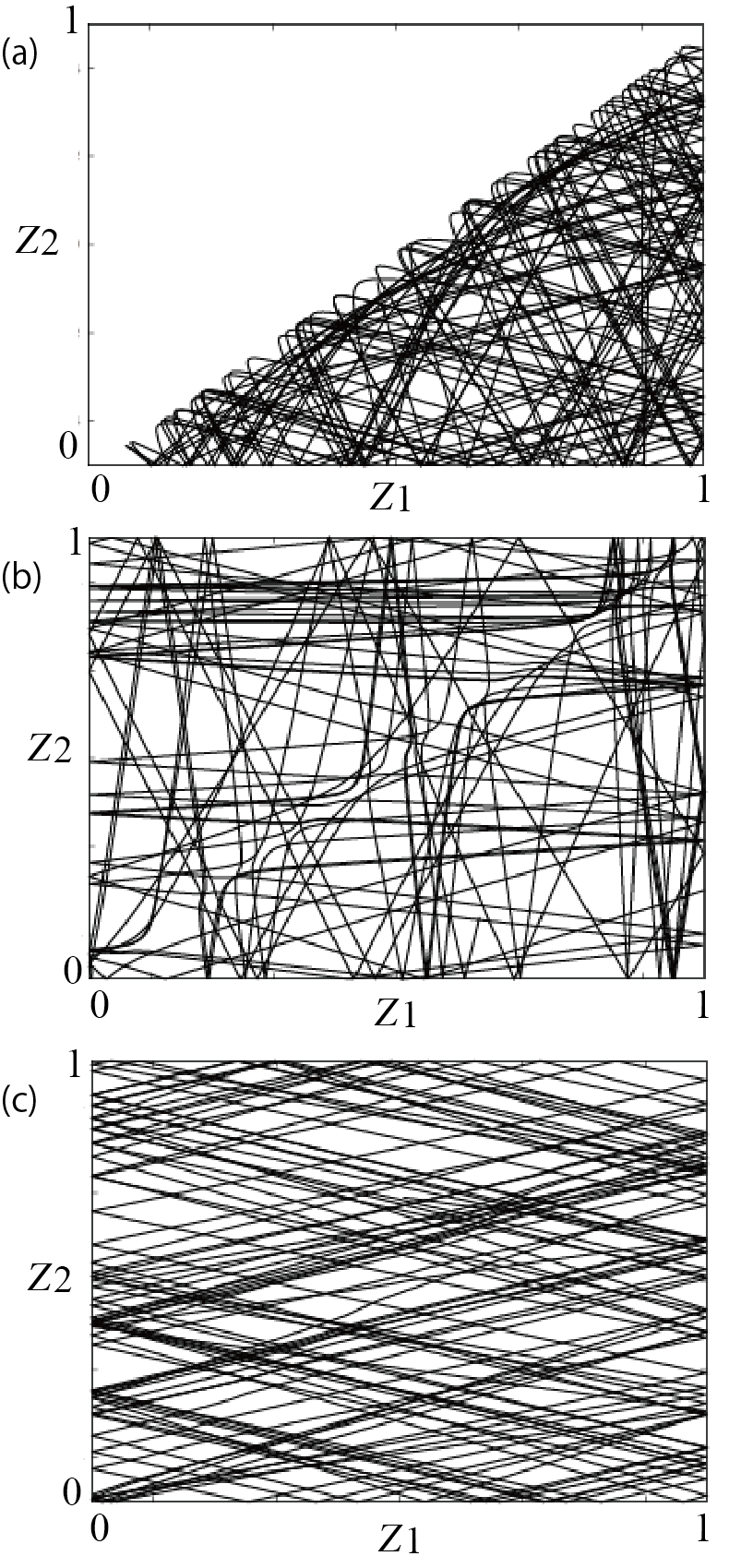}% Here is how to import EPS art
\end{center}
\caption{\label{fig:epsart}
Typical trajectories $(z_1(t),z_2(t))$ with (a) $\ep=50$,(b) $\ep=200$,
(c) $\ep=1000$, respectively.
}
\label{orbits}
\end{figure}

In Fig.\ref{orbits} typical trajectories $(z_1(t),z_2(t))$ are shown for 
(a) $\ep=50$,(b) $\ep=200$ and (c) $\ep=1000$, respectively.
The Poincar\'e maps for those orbits show that they are chaotic.
It is seen that the potential bends the trajectories 
especially near $z_2=z_1$ line for $\ep=50$ and $\ep=200$.
It causes irregularity on the orbits.
Contrastively the effect of potential is much less for $\ep=1000$.
The trajectory is composed of nearly straight lines.
The orbits become more regular 
for larger $\ep$ if $\ep$ is large enough.

\begin{figure}[h!]
\begin{center}
\includegraphics[width=7cm]{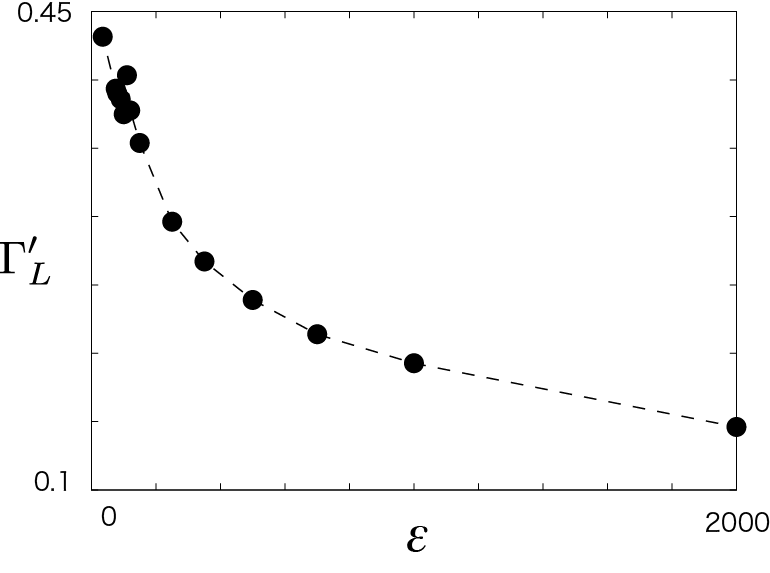}% Here is how to import EPS art
\end{center}
\caption{\label{fig:epsart}
$\ep-$dependence of $\Gamma_L'$ in Eq.(\ref{lia5}).
The broken line is a guide to the eyes.
}
\label{lia_2_store_seki}
\end{figure}

In order to confirm this point quantitatively we evaluate
$\Gamma'_L$ in Eq.(\ref{lia5}) with $\xi=0.01$, which reflects irregularity of
trajectories.
%$\xi$ in Eq.(\ref{lia5}) is chosen to be equal to $0.01$.
Numerical calculation of each trajectory is performed 
for time more than $1.5\times 10^5$. 
We take the average of $\Gamma'_L$ over about 20 orbits corresponding to the
largest irregular region in the Poincar\'e maps for each $\ep$.
We show the $\ep-$dependence of $\Gamma'_L$ in Fig.\ref{lia_2_store_seki}.
The decrease of $\Gamma'_L$ indicates the fact that the dynamics becomes
more regular with the increase of $\ep$, 
which is consistent with the above intuitive view from Fig.\ref{orbits}. 
It is also consistent with the $\ep-$dependence of $\alpha$ in the quantum
system.

\begin{figure}[h!]
\begin{center}
\includegraphics[width=7cm]{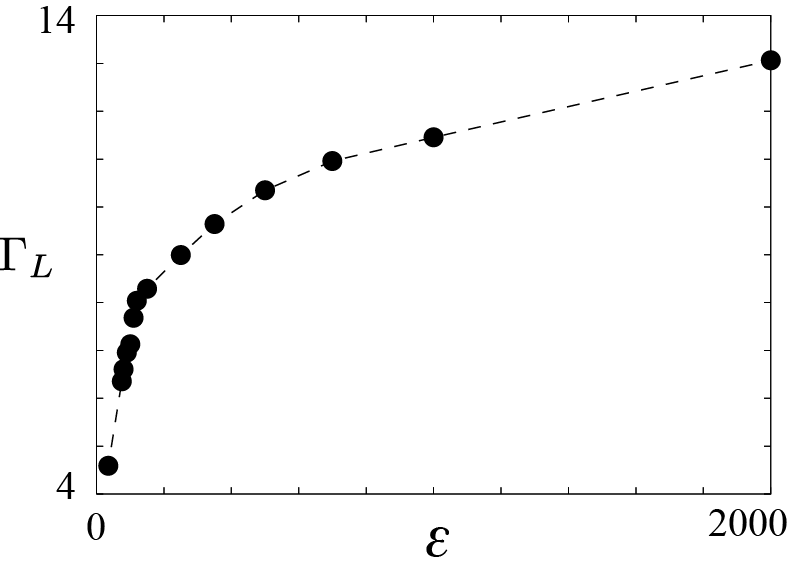}% Here is how to import EPS art
\end{center}
\caption{\label{fig:epsart}The same as Fig.\ref{lia_2_store_seki} except for 
$\Gamma_L$ in Eq.(\ref{lia4}).
}
\label{lia_1_store_seki}
\end{figure}

Now we calculate the ordinary MLE 
$\Gamma_L$ in Eq.(\ref{lia4}) with $\Delta\tau = 0.01$.
Numerical calculations for trajectories are performed 
for time more than $1.5\times 10^5$.  
We take the average of $\Gamma_L$ over about 20 orbits corresponding to the
largest irregular region in the Poincar\'e maps as well as for $\Gamma_L'$.
The $\ep-$dependence of $\Gamma_L$ is shown in Fig.\ref{lia_1_store_seki}.
We see that $\Gamma_L$ increases with $\ep$. 
This is because motions of the particles become faster 
with the increase of $\ep$
and does not necessarily imply the increase of chaotic irregularity.
Therefore $\ep-$dependence of $\Gamma_L$ does not directly 
reflect the degree of 
chaotic irregularity.

\begin{figure}[h!]
\begin{center}
\includegraphics[width=7cm]{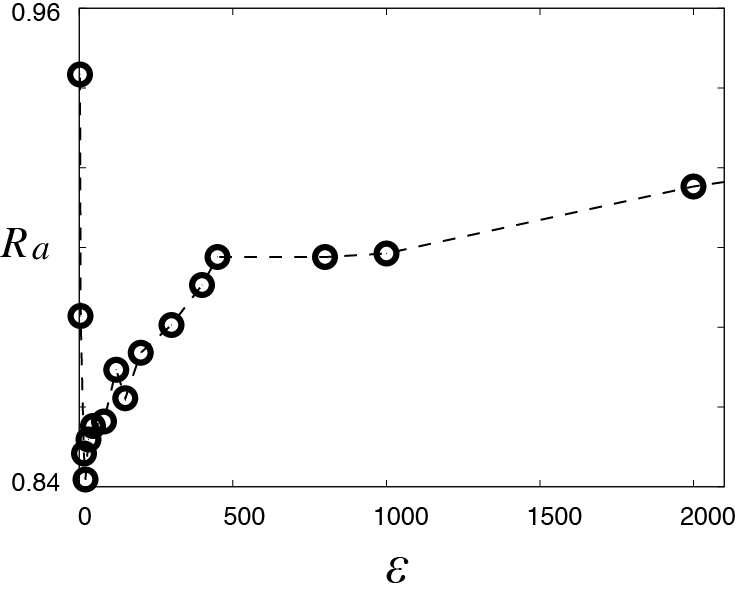}% Here is how to import EPS art
\end{center}
\caption{\label{fig:epsart}
$\ep-$dependence of the ratio $R_a$ between two areas in Poincar\'e map 1:
the area of the largest irregular region and area of total 
region reachable for a particle with $\ep$.
The broken line is a guide to the eyes.
}
\label{area_map1_en}
\end{figure}

\begin{figure}[h!]
\begin{center}
\includegraphics[width=7cm]{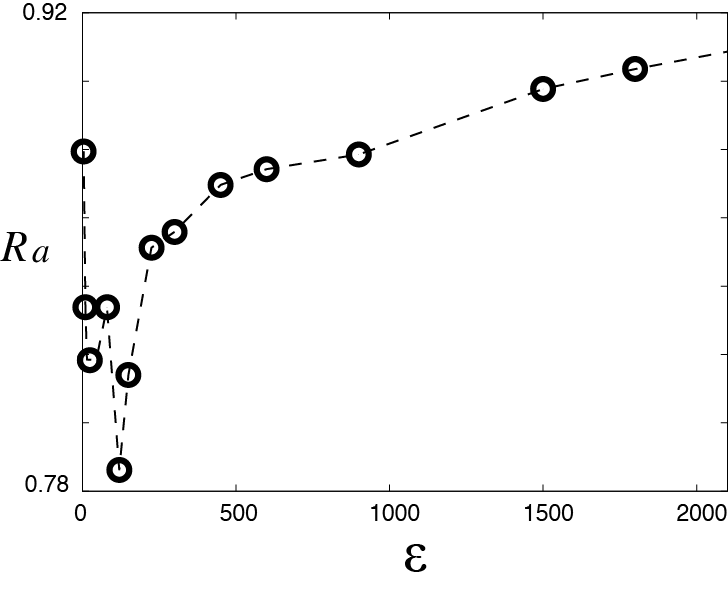}% Here is how to import EPS art
\end{center}
\caption{\label{fig:epsart}
The same as Fig.\ref{area_map1_en} except for Poincar\'e map 2.
}
\label{area_map2_en}
\end{figure}

We also point out that an area of the largest irregular 
region in Poincar\'e map, 
which is adopted by several authors as a measure of chaotic
irregularity \cite{Win,Ter,Har}, is irrelevant for the present system. 
%As seen in Figs.\ref{map_E50_jpg}
%-\ref{map_E1000_Bro}, 
%the area of the irregular regions look larger for larger $\ep$.
We calculate the ratio $R_a$ between two areas in Poincar\'e maps:
the area of the largest irregular region and the area of total 
region reachable for a particle with $\ep$. 
For the calculations we take meshes on Poincar\'e map.
A total number of meshes is $300\times 300$.
Then we count the 
number of meshes which an irregular trajectory visits 
and compare it to the total number of meshes  energetically allowed.
Numerical calculations of the trajectories are performed 
for time more than $2.0\times 10^5$.
The $\ep-$dependence of $R_a$ for Poincar\'e map 1 and 2 are shown
in Figs.\ref{area_map1_en} and \ref{area_map2_en}, respectively.
It is seen that $R_a$ increases with $\ep$ for $\ep>100$,
in which the trajectory in two-dimensional square can cross
the potential hill (the diagonal line).
%As shown in Fig.\ref{lia_2_store_seki}, $\Gamma_L'$ which indicates the 
%intensity of irregularity decreases with $\ep$.
%While the orbit becomes more regular when $\ep$ increases
%as seen obviously in Fig.\ref{orbits},
%$R_a$ can be large even in the system with very weak irregularity.
On the other hand, the orbits become more regular as seen obviously in 
Fig.\ref{orbits} when $\ep$ increases for $\ep>100$.
Therefore $R_a$ is not a proper measure of irregularity 
in the present system in contrast to the other systems, in which $R_a$ can be 
adopted as a measure of irregularity \cite{Win,Ter,Har}.

\clearpage
\subsection{Attractive interaction}
\begin{figure}[h!]
\begin{center}
\includegraphics[width=7cm]{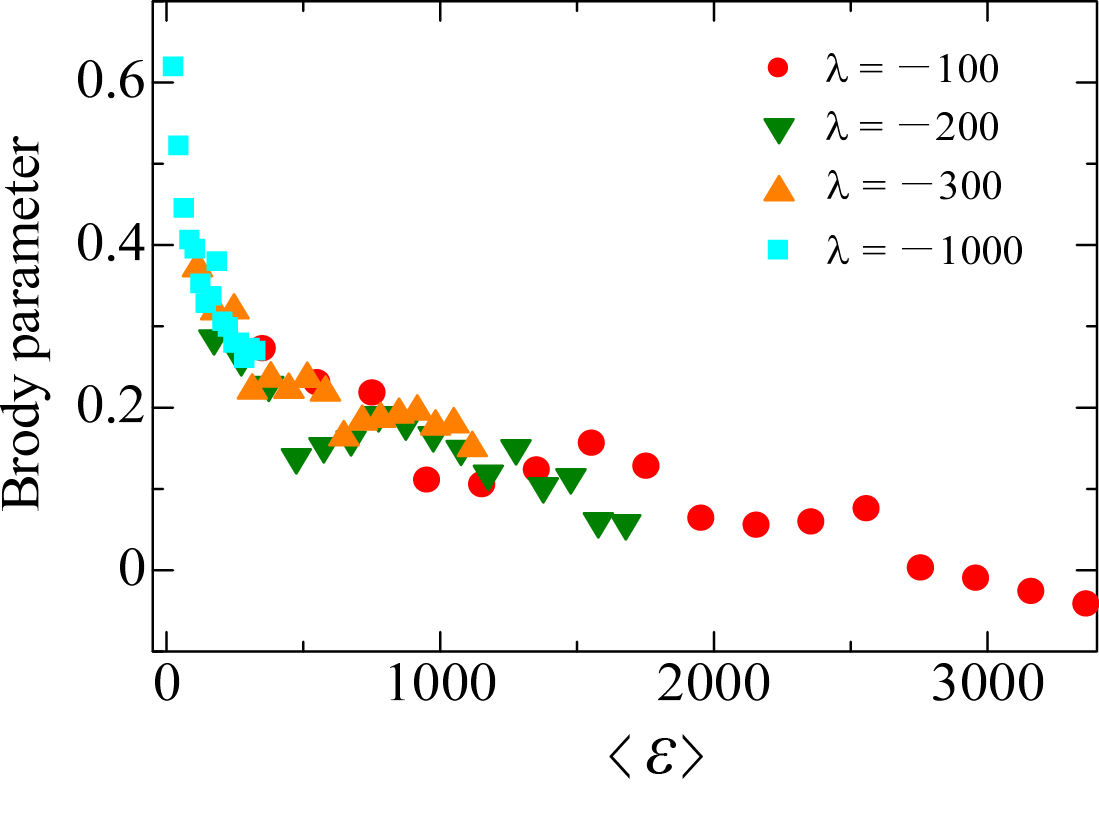}% Here is how to import EPS art
\end{center}
\caption{\label{fig:epsart}
$\ep-$dependence of the Brody parameter for 
the attractive interaction.
For each data point eigenvalues in the energy interval $\Delta=50000$ 
were used (
The number of eigenvalues used to calculate each data point is about 1000).
Horizontal axis denotes the average of $\ep$ of the used eigenstates.
}
\label{bro_store_inryoku}
\end{figure}

\begin{figure}[h!]
\begin{center}
\includegraphics[width=7cm]{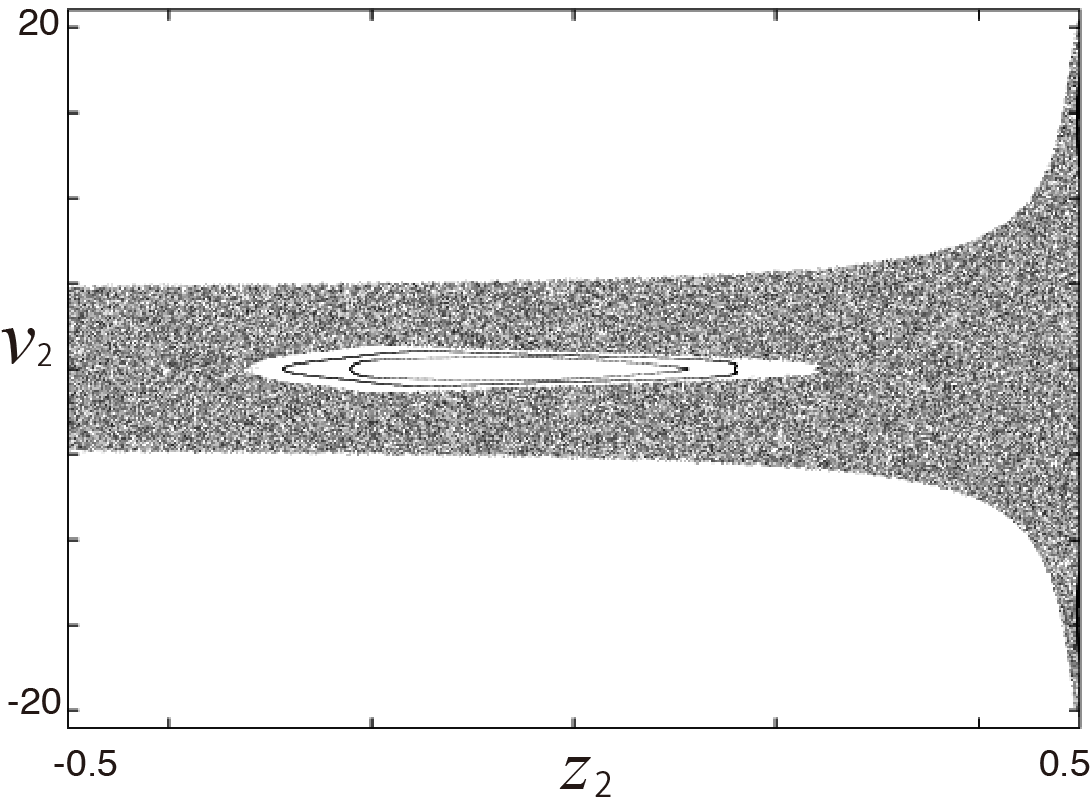}% Here is how to import EPS art
\end{center}
\caption{\label{fig:epsart}
Poincar\'e map 1 for attractive interaction with $\ep=5$.
}
\label{map_E5_inryoku_yamada_pic}
\end{figure}

\begin{figure}[h!]
\begin{center}
\includegraphics[width=7cm]{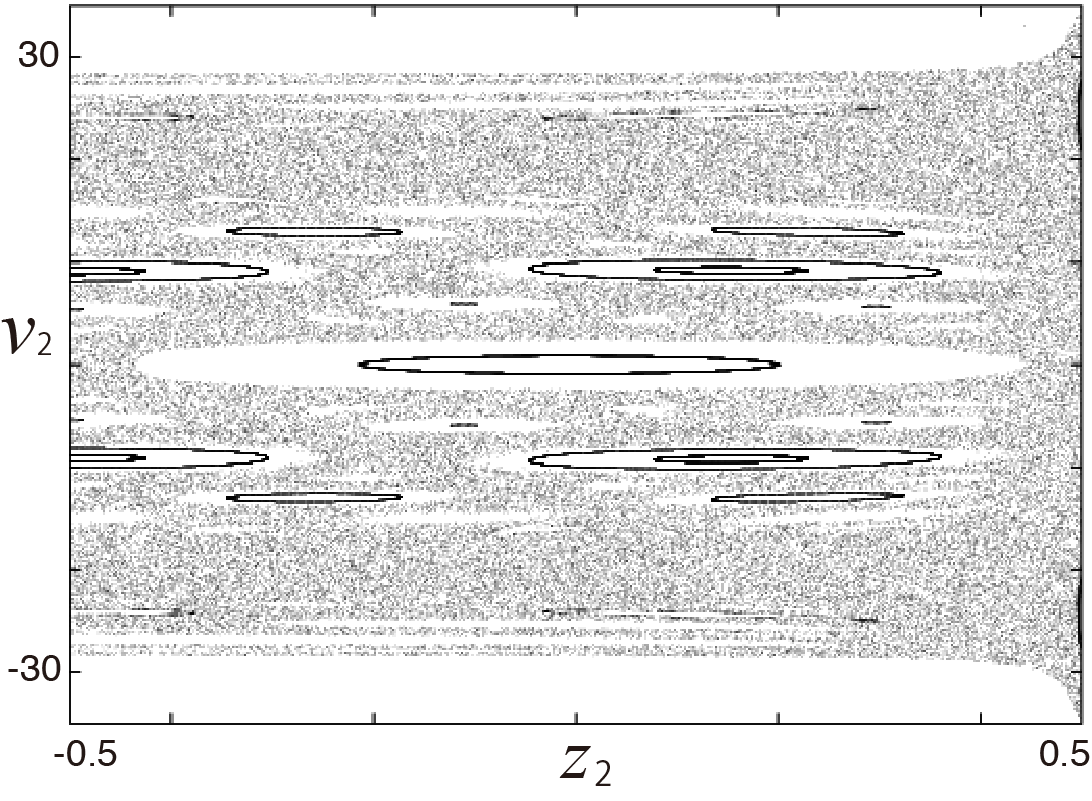}% Here is how to import EPS art
\end{center}
\caption{\label{fig:epsart}
Poincar\'e map 1 for attractive interaction with $\ep=200$.
}
\label{map_E200_inryoku_yamada_pic}
\end{figure}

\begin{figure}[h!]
\begin{center}
\includegraphics[width=7cm]{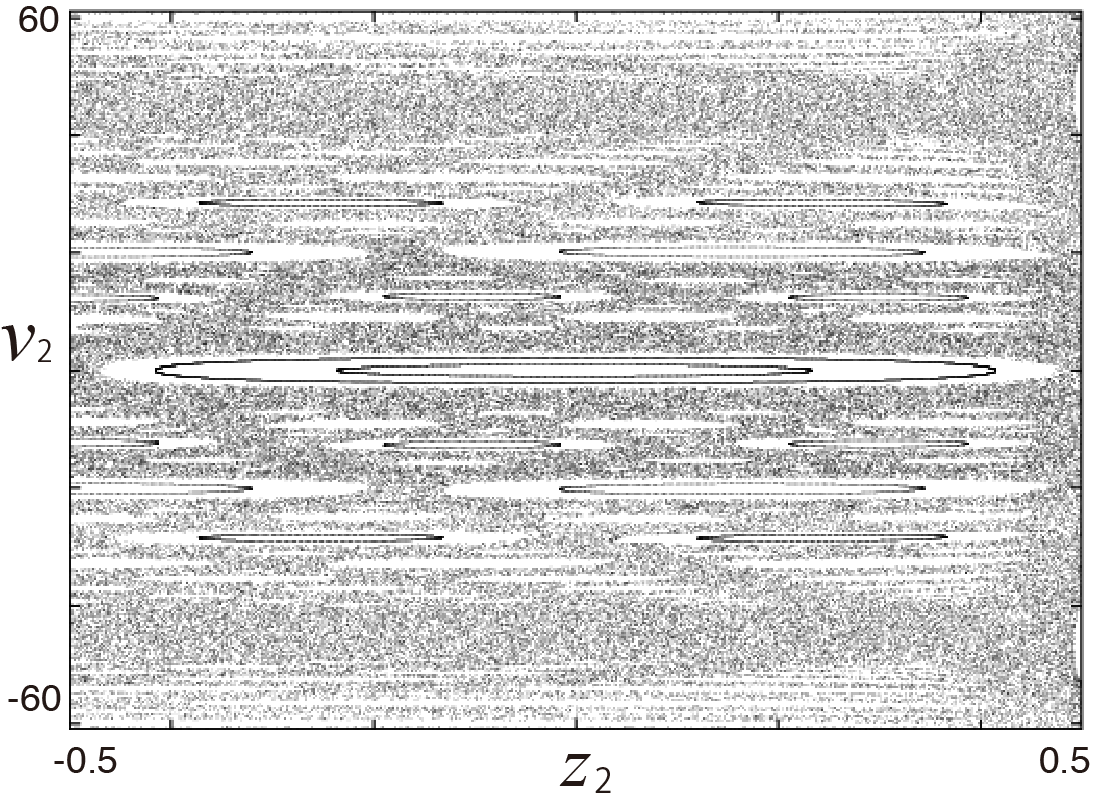}% Here is how to import EPS art
\end{center}
\caption{\label{fig:epsart}
Poincar\'e map 1 for attractive interaction with $\ep=1000$.
}
\label{map_E1000_inryoku_yamada_pic}
\end{figure}

\begin{figure}[h!]
\begin{center}
\includegraphics[width=7cm]{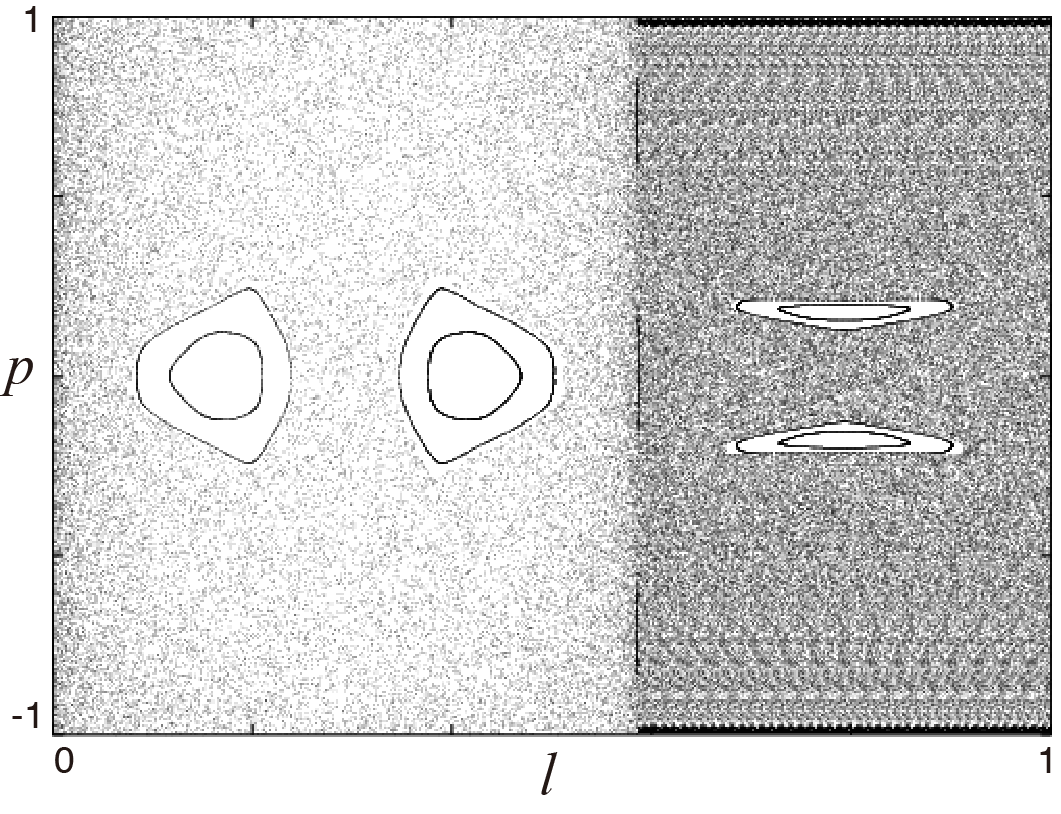}% Here is how to import EPS art
\end{center}
\caption{\label{fig:epsart}
Poincar\'e map 2 for attractive interaction with $\ep=5$.
}
\label{map_E5_Bro_inryoku}
\end{figure}

\begin{figure}[h!]
\begin{center}
\includegraphics[width=7cm]{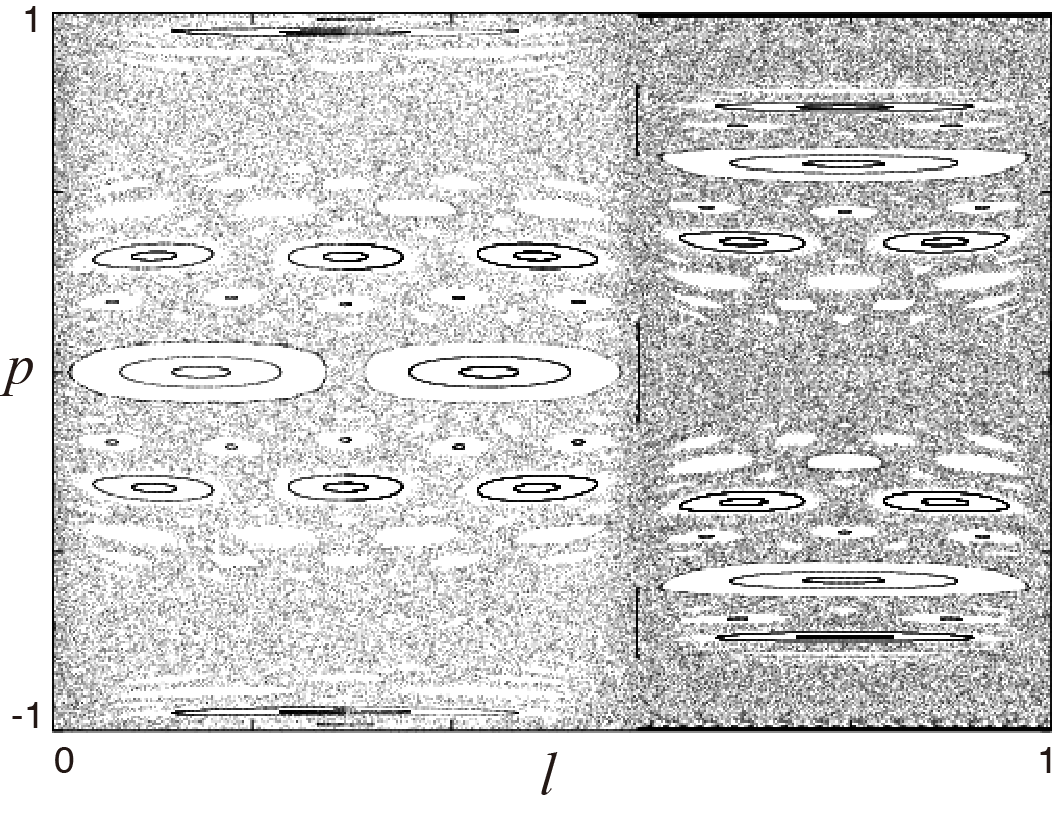}% Here is how to import EPS art
\end{center}
\caption{\label{fig:epsart}
Poincar\'e map 2 for attractive interaction with $\ep=200$.
}
\label{map_E200_Bro_inryoku}
\end{figure}

\begin{figure}[h!]
\begin{center}
\includegraphics[width=7cm]{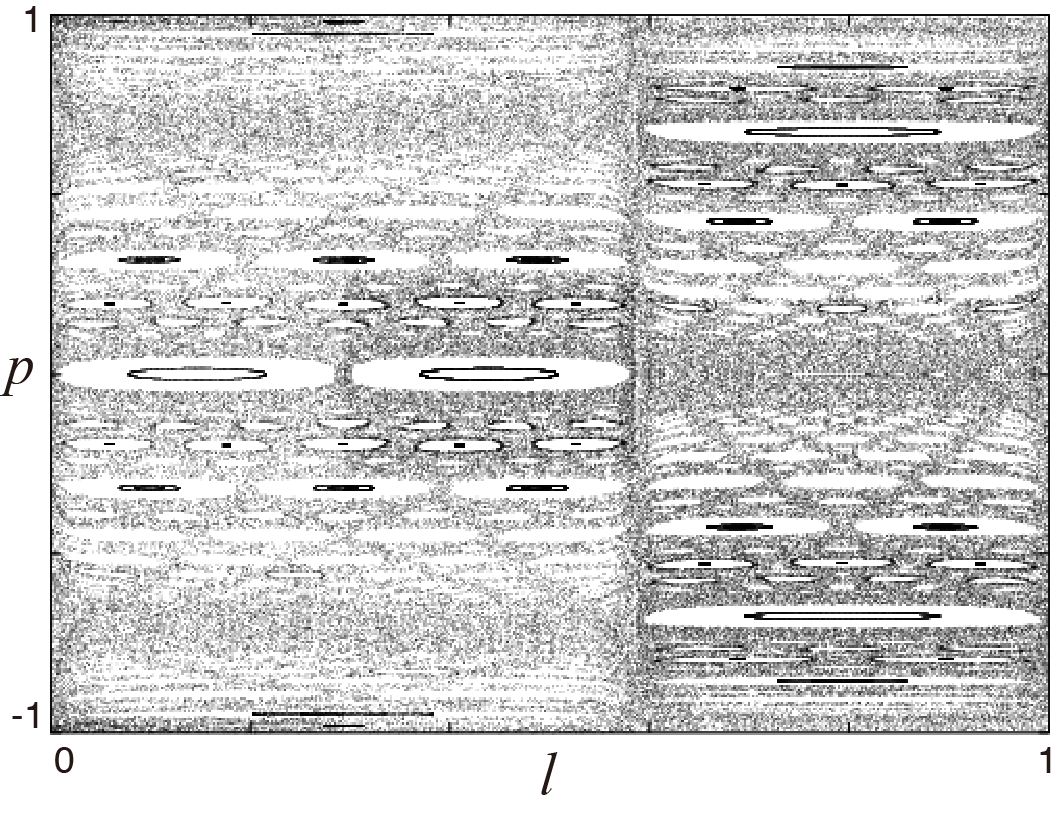}% Here is how to import EPS art
\end{center}
\caption{\label{fig:epsart}
Poincar\'e map 2 for attractive interaction with $\ep=1000$.
}
\label{map_E1000_Bro_inryoku}
\end{figure}

Now we turn to results for the case of the attractive 
interaction with $\lam < 0$ and $\delta=0.01$.
The $\ep-$dependence of the Brody parameter
is shown in Fig.\ref{bro_store_inryoku}.
$<\ep>$ is the average of $\ep$ of the used eigenstates.
The decrease of the Brody parameter is seen with the increase of 
the scaled energy $\ep$.
We see that the Brody parameter depends almost only on $\ep$
and not on $E$ and $\lam$ separately, which is a situation similar to 
the case of the repulsive interaction.
%The $\ep-$dependences of the Brody parameter for different values of $\lam$ 
%are similar to each other.
Next,
we consider the corresponding classical dynamics described by
the equations of motion in Eq.(\ref{eq7_7_1}) for the attractive interaction.
%The energy is greater than $-1/\delta$.
The Poincar\'e maps 1 are shown 
for $\ep=5, 200, 1000$ 
in Figs.\ref{map_E5_inryoku_yamada_pic} - \ref{map_E1000_inryoku_yamada_pic},
respectively.
The Poincar\'e maps 2 are also shown 
in Figs.\ref{map_E5_Bro_inryoku} - \ref{map_E1000_Bro_inryoku}.
We have taken about 20 different initial points in phase space for each map.
These Poincar\'{e} maps show that the present classical system
exhibits mixed dynamics with coexisting KAM tri and chaotic regions,
as well as in the case of repulsive interaction.

\begin{figure}[h!]
\begin{center}
\includegraphics[width=7cm]{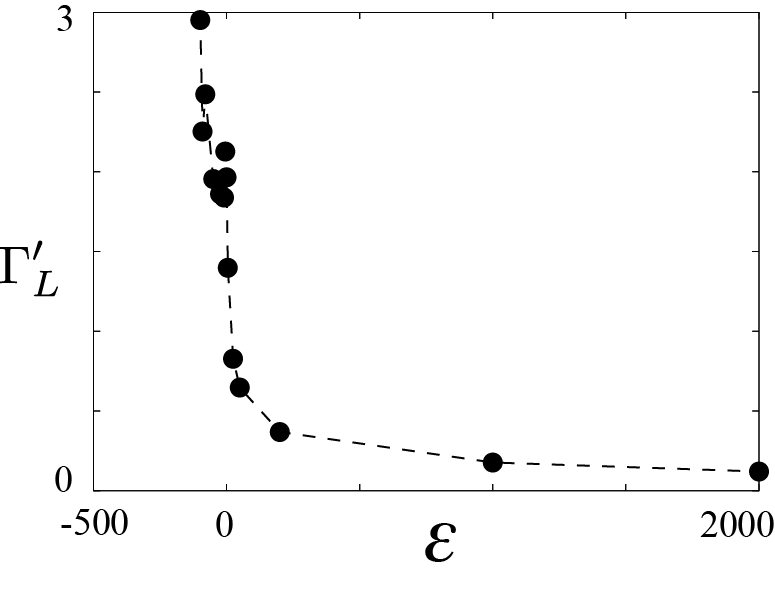}% Here is how to import EPS art
\end{center}
\caption{\label{fig:epsart}$\ep-$dependence of
$\Gamma_L'$ in Eq.(\ref{lia5}) for the attractive interaction.
The broken line is a guide to the eyes.
}
\label{lia_3_store_at}
\end{figure}

We calculate $\Gamma_L'$ 
in Eq.(\ref{lia5}) with $\xi=0.01$, which reflects the 
degree of chaotic irregularity.
Numerical calculations of the trajectories are performed 
for time more than $1.5\times 10^5$.
The results are shown in Fig.\ref{lia_3_store_at}, where we see that 
$\Gamma_L'$ decreases with increase of $\ep$ for $\ep>0$.
%For $\ep>0$ the particle bounce on the square hard wall.
This is consistent with the fact that the dynamics of the particle 
becomes regular when $\ep$ increases for $\ep>0$.
\begin{figure}[h!]
\begin{center}
\includegraphics[width=7cm]{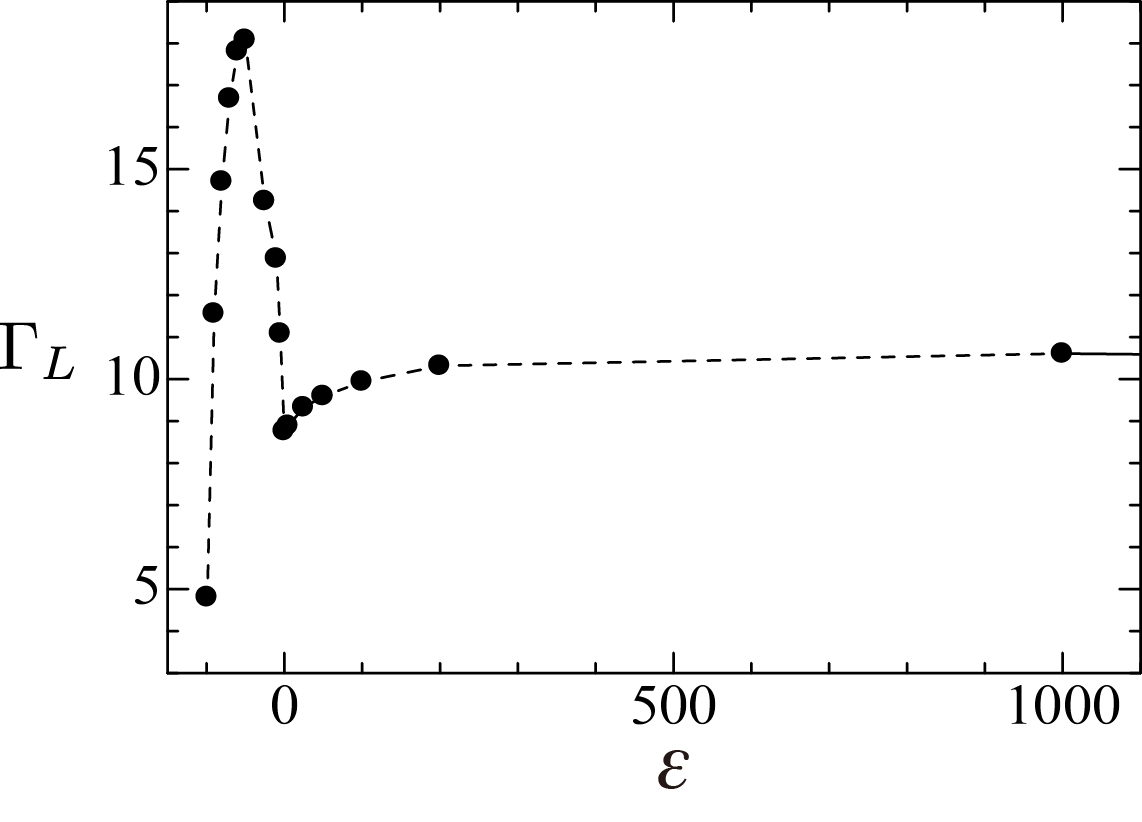}% Here is how to import EPS art
\end{center}
\caption{\label{fig:epsart}
The same as Fig.\ref{lia_3_store_at} except for MLE $\Gamma_L$ 
in Eq.(\ref{lia4}).
}
\label{lia_2_store_at}
\end{figure}
On the other hand, $\Gamma_L$ in Eq.(\ref{lia4}) 
increases with $\ep$ as seen in Fig.\ref{lia_2_store_at},
where we take $\Delta t=0.01$.
The increase of $\Gamma_L$ is due to the fact that
the dynamics of the particle becomes faster with the increase of $\ep$.
$\Gamma_L$ does not reflect degree of chaotic irregularity
of the classical system similarly to the case of the repulsive interaction.

\begin{figure}[h!]
\begin{center}
\includegraphics[width=7cm]{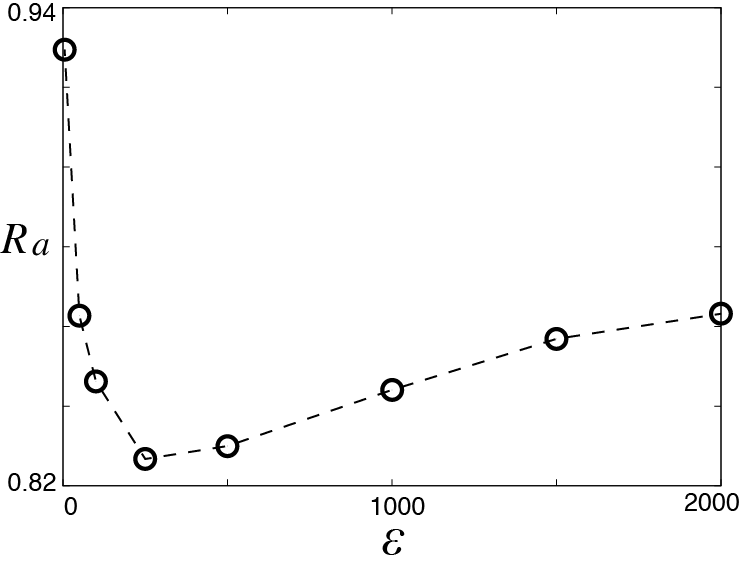}% Here is how to import EPS art
\end{center}
\caption{\label{fig:epsart}
$\ep-$dependence of the ratio $R_a$ between two areas in Poincar\'e map 1.
The broken line is a guide to the eyes.
}
\label{e_area_sawada_inryoku}
\end{figure}

\begin{figure}[h!]
\begin{center}
\includegraphics[width=7cm]{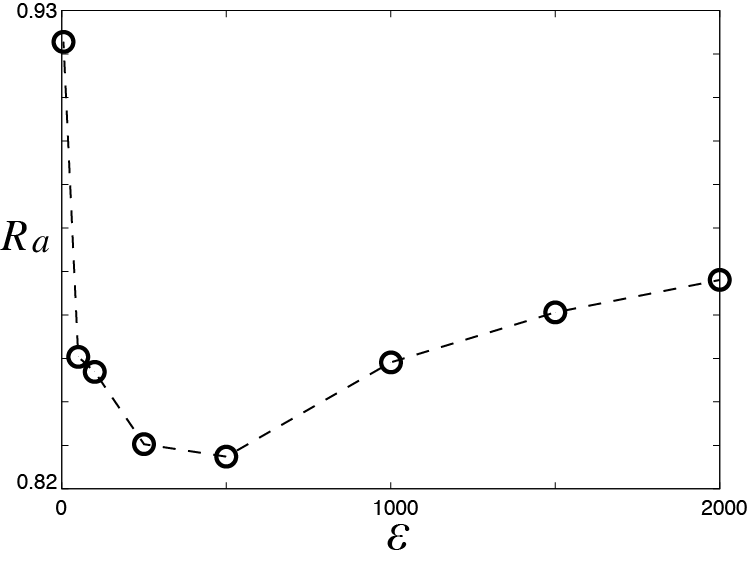}% Here is how to import EPS art
\end{center}
\caption{\label{fig:epsart}
The same as Fig.\ref{e_area_sawada_inryoku} except for Poincar\'e map 2.
}
\label{e_area_shimizu_inryoku}
\end{figure}

We also calculate the
ratio $R_a$ between two areas in Poincar\'e maps,
the area of the largest irregular region and the area of total 
region reachable, in the same manner used for 
repulsive interaction. 
The $\ep-$dependence of $R_a$ for Poincar\'e map 1 and 2 are shown
in Figs.\ref{e_area_sawada_inryoku} and 
\ref{e_area_shimizu_inryoku}, respectively.
$R_a$ increases with $\ep$ for $\ep>500$,
while the orbit becomes more regular when $\ep$ increases
as well as in the case of the repulsive interaction.
%$R_a$ can be large even in the system with very weak irregularity.
Therefore $R_a$ is not a proper measure of irregularity also for
the case of the attractive interaction in present system.

\clearpage

\section{summary and conclusion}
\label{conclusion}
We studied dynamics of two spinless particles confined in a quantum wire
with repulsive or attractive Coulomb interaction.
The system is reduced to a quasi-one-dimensional system with effective potential
under the assumption that the transverse confinement is much stronger than the
longitudinal one.

The Coulomb interaction induces irregular dynamics in classical mechanics.
Examining Poincar\'e maps for the present system,
we have found that the classical system exhibits mixed dynamics with coexisting 
KAM tori and chaotic regions.
To see the signatures of quantum chaos in the corresponding quantum 
system we analyzed the distributions of the nearest neighbor level spacing
(NNLS), which is fitted to the Brody
distribution function characterized by the Brody parameter $\alpha$.
The results indicate that they are intermediate between the Poisson and Wigner
distributions, which is consistent with the mixed character of the classical 
dynamics.

The present classical system has a scaling property: Its dynamics is
characterized by the rescaled energy parameter $\ep=E/|\lam|$, where $\lam$
is the interaction strength parameter. Contrastingly, 
the quantum system has no such scaling property. However it has turned out
that the distribution of NNLS in the quantum system has a scaling property
similarly to the case of classical mechanics. The Brody parameter $\alpha$
depends almost only on the average value of $\ep$ and is insensitive to the
value of $\lam$ itself. 

In the classical system, 
we found that orbits are more regular
for larger values of $\ep$.
The ordinary MLE $\Gamma_L$ is not 
suitable measure of chaotic irregularity for the present system,
because they increase with $\ep$ whereas
the irregularity of the system decreases.
We introduced a new MLE $\Gamma_L'$, 
which represents a rate of the exponential divergence of two adjacent orbits 
(reference and displaced orbits) with respect to the length of the reference 
orbit, while the ordinary Lyapunov exponent describes the one with respect to 
time.
The dependence of $\Gamma_L'$ on $\ep$ quantitatively shows 
the decrease of chaotic irregularity with increase of $\ep$.
Therefore, the $\Gamma_L'$ is a suitable measure of chaotic irregularity of the 
present classical system rather than $\Gamma_L$.
On the other hand in quantum system, the Brody parameter $\alpha$ decreases
almost monotonously with increase of the average value of $\ep$ in both
cases of the repulsive and attractive interactions which indicates
the distribution function of NNLS approaches to Poisson distribution
with increase of $\ep$.
Consequently, we have shown closer correspondence between the classical 
chaos and quantum chaos in the present system. 

We also showed that
the area of the irregular region in Poincar\'e maps are not 
suitable measure of chaotic irregularity for the present system
in contrast with other systems in which the area has been
adopted as a measure of irregularity by several authors \cite{Win,Ter,Har}.

\begin{acknowledgements}
One of the authors (S. M.) thanks global COE program 
``Weaving Science Web beyond Particle-Matter Hierarchy'' for its 
financial support.
This work is partly supported by JSPS KAKENHI(Grant No.23540459).
\end{acknowledgements}

\end{document}